\documentclass[twocolumn]{aastex631}

\shorttitle{Cosmic Evolution of \ion{Fe}{2}/\ion{Mg}{2} in Quasar BLRs}
\shortauthors{Jiang et al.}

\begin{document}
\title{No Redshift Evolution in the \ion{Fe}{2}/\ion{Mg}{2} Flux Ratios of Quasars across Cosmic Time}

\author[0009-0003-6747-2221]{Danyang Jiang}
\affiliation{Department of Astronomy, School of Physics, Peking University, Beijing 100871, China}
\affiliation{Kavli Institute for Astronomy and Astrophysics, Peking University, Beijing 100871, China}

\author[0000-0003-2984-6803]{Masafusa Onoue}
\altaffiliation{Kavli Astrophysics Fellow}
\affiliation{Kavli Institute for Astronomy and Astrophysics, Peking University, Beijing 100871, China}
\affiliation{Kavli Institute for the Physics and Mathematics of the
Universe (Kavli IPMU, WPI), The University of Tokyo Institutes for Advanced Study, The University of Tokyo, Kashiwa, Chiba 277-8583, Japan}
\affiliation{Center for Data-Driven Discovery, Kavli IPMU (WPI), UTIAS, The University of Tokyo, Kashiwa, Chiba 277-8583, Japan}

\author[0000-0003-4176-6486]{Linhua Jiang}
\affiliation{Department of Astronomy, School of Physics, Peking University, Beijing 100871, China}
\affiliation{Kavli Institute for Astronomy and Astrophysics, Peking University, Beijing 100871, China}

\author[0000-0001-9372-4611]{Samuel Lai}
\affiliation{Research School of Astronomy and Astrophysics, Australian National University, Canberra, ACT 2611, Australia}

\author[0000-0002-2931-7824]{Eduardo Ba{\~n}ados}
\affiliation{Max Planck Institut f\"ur Astronomie, K\"onigstuhl 17, D-69117, Heidelberg, Germany}

\author[0000-0003-2344-263X]{George D. Becker}
\affiliation{Department of Physics and Astronomy, University of California, Riverside, CA 92521, USA}

\author[0000-0003-2344-263X]{Manuela Bischetti}
\affiliation{Dipartimento di Fisica, Sezione di Astronomia, Universitá di Trieste, via Tiepolo 11, I-34143 Trieste, Italy}

\author[0000-0001-8582-7012]{Sarah E. I. Bosman}
\affiliation{Institute for Theoretical Physics, Heidelberg University, Philosophenweg 12, D-69120, Heidelberg, Germany}
\affiliation{Max Planck Institut f\"ur Astronomie, K\"onigstuhl 17, D-69117, Heidelberg, Germany}

\author[0000-0002-3324-4824]{Rebecca L. Davies}
\affiliation{Centre for Astrophysics and Supercomputing, Swinburne University of Technology, Hawthorn, Victoria 3122, Australia}
\affiliation{ARC Centre of Excellence for All Sky Astrophysics in 3 Dimensions (ASTRO 3D), Australia}

\author[0000-0003-3693-3091]{Valentina D’Odorico}
\affiliation{INAF–Osservatorio Astronomico di Trieste, Via G.B. Tiepolo, 11, I-34143 Trieste, Italy}
\affiliation{Scuola Normale Superiore, P.zza dei Cavalieri, I-56126 Pisa, Italy}
\affiliation{IFPU–Institute for Fundamental Physics of the Universe, via Beirut 2, I-34151 Trieste, Italy}

\author[0000-0002-6822-2254]{Emanuele Paolo Farina}
\affiliation{Gemini Observatory, NSF’s NOIRLab, 670 N A’ohoku Place, Hilo, Hawai'i 96720, USA}

\author[0000-0001-8443-2393]{Martin G. Haehnelt}
\affiliation{Institute of Astronomy and Kavli Institute of Cosmology, University of Cambridge, Madingley Road, Cambridge CB3 0HA, UK}

\author[0000-0002-5941-5214]{Chiara Mazzucchelli}
\affiliation{Instituto de Estudios Astrof\'{\i}sicos, Facultad de Ingenier\'{\i}a y Ciencias, Universidad Diego Portales, Avenida Ejercito Libertador 441, Santiago, Chile.}

\author[0000-0002-4544-8242]{Jan-Torge Schindler}
\affiliation{Hamburger Sternwarte, Universit\"{a}t Hamburg, Gojenbergsweg 112, D-21029 Hamburg, Germany}

\author[0000-0003-4793-7880]{Fabian Walter}
\affiliation{Max Planck Institut f\"ur Astronomie, K\"onigstuhl 17, D-69117, Heidelberg, Germany}

\author[0000-0003-3307-7525]{Yongda Zhu}
\affiliation{Department of Physics and Astronomy, University of California, Riverside, CA 92521, USA}

\begin{abstract}

The \ion{Fe}{2}/\ion{Mg}{2} emission line flux ratio in quasar spectra serves as a proxy for the relative Fe to $\alpha$-element abundances in the broad line regions of quasars. Due to the expected different enrichment timescales of the two elements, they can be used as a cosmic clock in the early Universe. We present a study of the \ion{Fe}{2}/\ion{Mg}{2} ratios in a sample of luminous quasars exploiting high-quality near-IR spectra taken primarily by the XQR-30 program with VLT XSHOOTER. These quasars have a median bolometric luminosity of log($L_\mathrm{bol}$[erg s$^{-1}$]) $\sim47.3$ and cover a redshift range of  $z=6.0$--$6.6$. The median value of the measured \ion{Fe}{2}/\ion{Mg}{2} ratios is $\sim7.9$ with a normalized median absolute deviation of $\sim2.2$. In order to trace the cosmic evolution of \ion{Fe}{2}/\ion{Mg}{2} in an unbiased manner, we select two comparison samples of quasars with similar luminosities and high-quality spectra from the literature, one at intermediate redshifts ($z=3.5$--$4.8$) and the other at low redshifts ($z=1.0$--$2.0$). 
We perform the same spectral analysis for all these quasars, including the usage of the same iron template, the same spectral fitting method, and the same wavelength fitting windows.
We find no significant redshift evolution in the \ion{Fe}{2}/\ion{Mg}{2} ratio over the wide redshift range from $z=1$ to 6.6. The result is consistent with previous studies and supports the scenario of a rapid iron enrichment in the vicinity of accreting supermassive black holes at high redshift.

\end{abstract}


\section{Introduction} \label{sec:introduction}

The first-generation galaxies and quasars formed and lit up the dark Universe with ionizing photons when the age of the Universe was less than 1 Gyr. High-redshift ($z\ga6$) quasars provide a direct probe of this epoch of cosmic reionization thanks to their high luminosities \citep[e.g.,][]{Becker2015, Bosman2022}. In the past two decades, a number of quasars at $z \sim 6–7.5$ have been discovered \citep[e.g.,][]{2001ApJ...122...2833, 2007AJ...134...1150,2016ApJ...833...222J, 2010ApJ...139...906, 2013ApJ...779...24, 2015MNRAS.454.3952R, Banados2016, Banados2018, 2018ApJS...237...5, 2022ApJS..259...18M, 2019ApJ...884...30W, 2021ApJ...907...L1}. Most of those high-redshift luminous quasars host supermassive black holes (SMBHs) with BH masses $M_{\rm BH}\sim10^{9-10} M_{\sun}$, accreting near the Eddington limit \citep[e.g.,][]{farina2022, mazzucchelli2023}. 
The dense gas orbiting around SMBHs in the broad line region (BLR) is observed as broad emission lines with the typical full width at half maximum (FWHM) $\approx 2000$--$10,000$ km s$^{-1}$ at rest-frame UV and optical wavelengths. Photoionization models suggest that the flux ratios of broad emission lines such as \ion{N}{5}/\ion{C}{4}, \ion{N}{5}/\ion{He}{2} and (\ion{Si}{4}+\ion{O}{4})/\ion{C}{4} can be used to estimate the BLR metallicity \citep[][but also see \citealt{Temple21} for the influence of non-abundance parameters]{2002ApJ...564...592, 2006A&A...447...1}. While the BLR metallicity is expected to reflect the star formation history of the quasar host galaxy, current observations suggest no significant redshift evolution of the metallicity at redshift up to $z\sim7.5$ \citep[e.g.,][]{2006A&A...447...1, 2007AJ...134...1150, 2009A&A...494...L25-L28, 2011ApJ...739...56, 2014ApJ...790...145, 2017ApJ...849...91, 2019MNRAS...484...2575, 2020ApJ...898...105,2020ApJ...905...51, 2022MNRAS...513...1801--1819, 2022ApJ...925...121}.

Among all BLR line flux ratios, the ratio of the \ion{Fe}{2} pseudo-continuum in the rest-frame UV to the \ion{Mg}{2} doublet at rest-frame 2798 \AA\ (\ion{Fe}{2}/\ion{Mg}{2}~$\lambda$2798) is of particular interest. Mg, an $\alpha$-element, is enriched mainly by core-collapse type II supernovae (SNe II) soon after the onset of star formation \citep[e.g.,][]{woosley1992}. 
Fe is thought to be efficiently produced by type Ia supernovae (SNe Ia) that evolve from binary systems with a longer lifetime $\sim1$ Gyr \citep[e.g.,][]{1983A&A...118..217G}. The enrichment timescale of Fe is thus delayed by $\sim1$ Gyr relative to that of Mg. This time delay of the Fe enrichment can serve as a cosmic clock  \citep[e.g.,][]{1993ApJ...418...11H, 1999ARAA...37...487}. 
Given that $z\ge6$ quasars have reached the first billion years of the Universe, one expects to see a rapid decline of their \ion{Fe}{2}/\ion{Mg}{2} at $z>6$, which is sensitive to the star formation history in quasar host galaxies. Observationally, no obvious redshift evolution has been reported at redshift up to $z\sim7.5$ \citep[e.g.,][]{1996ApJ...470...L85, 1998ApJ...507...L113, 1999ApJ...515...487, 2002ApJ...564...581,2003ApJ....596...817,2002ApJ...565...63,2004ApJ...614...69, 2003ApJ...594...L95,2003ApJ...587...L67, 2003ApJ...596...L155, 2006ApJ...650...57, 2007AJ...134...1150, 2007ApJ...669...32, 2011ApJ...739...56,2014ApJ...790...145, 2017ApJ...834...203,2020ApJ...904...162, 2017ApJ...849...91, 2019ApJ...874...22,2021ApJ...917...107, 2020ApJ...898...105, 2020ApJ...905...51, Yang21, 2022ApJ...937...61,2022ApJ...925...121}. 
Based on the non-evolution of \ion{Fe}{2}/\ion{Mg}{2}, it has been suggested that there are other mechanisms for the Fe production, such as the presence of Population-III stars \citep{2020ApJ...898...105, 2022ApJ...937...61, 2022MNRAS...512...2573}.

In previous studies of the \ion{Fe}{2}/\ion{Mg}{2} redshift evolution, quasar samples and spectral analysis methods are often heterogeneous. 
Comparison samples at different redshifts were usually from different studies, which may introduce systematic uncertainties or even biases. For example, different studies usually used different spectral fitting methods and different iron emission templates \citep[e.g.,][]{2017ApJ...849...91,2019ApJ...874...22}. 
\citet{2020ApJ...898...105} showed that different templates of the UV \ion{Fe}{2} pseudo-continuum introduce an offset of 37 \%  in \ion{Fe}{2}/\ion{Mg}{2} \citep[see also e.g.,][]{woo2018, 2020ApJ...905...51}. 
In addition, different studies often used different wavelength ranges or spectral windows for emission line fitting \citep[e.g.,][]{2007AJ...134...1150,2017ApJ...834...203,2022ApJ...925...121}. This was limited by the spectra available for different redshifts. 
The impact of all above factors on the \ion{Fe}{2}/\ion{Mg}{2} measurement is not well understood. Therefore, it is critical to measure the \ion{Fe}{2}/\ion{Mg}{2} of quasars at a wide redshift range based on a unified approach.

This paper presents our \ion{Fe}{2}/\ion{Mg}{2} measurements of $6.0\lesssim z\lesssim 6.6$ quasars using high-quality spectra taken by a large spectroscopic campaign of VLT/XSHOOTER. The high-redshift sample is compared with intermediate-redshift ($z\sim4$) and low-redshift ($z\sim1.5$) samples in order to address the \ion{Fe}{2}/\ion{Mg}{2} evolution over cosmic time. 
The structure of the paper is as follows. We introduce our datasets and method of spectral analyses in Section \ref{sec:observation} and Section \ref{sec:data_analyses} respectively. The results of our \ion{Fe}{2}/\ion{Mg}{2} measurements  are shown in Section \ref{sec:results}. We discuss the systematic uncertainties and implications of our results in Section \ref{sec:discussion} and summarize the paper in Section \ref{sec:summary}. In this paper, we adopt a standard $\Lambda$CDM cosmology with $H_0$=70 km s$^{-1}$ Mpc$^{-1}$, $\Omega_\mathrm{m}$=0.3, and $\Omega_\mathrm{\Lambda}$=0.7.
All the magnitudes are reported in the AB system.

\section{Data and quasar sample} \label{sec:observation}

\subsection{The enlarged XQR-30 quasar sample} \label{subsec:data_XQR30}

\begin{figure*}[!]
\plotone{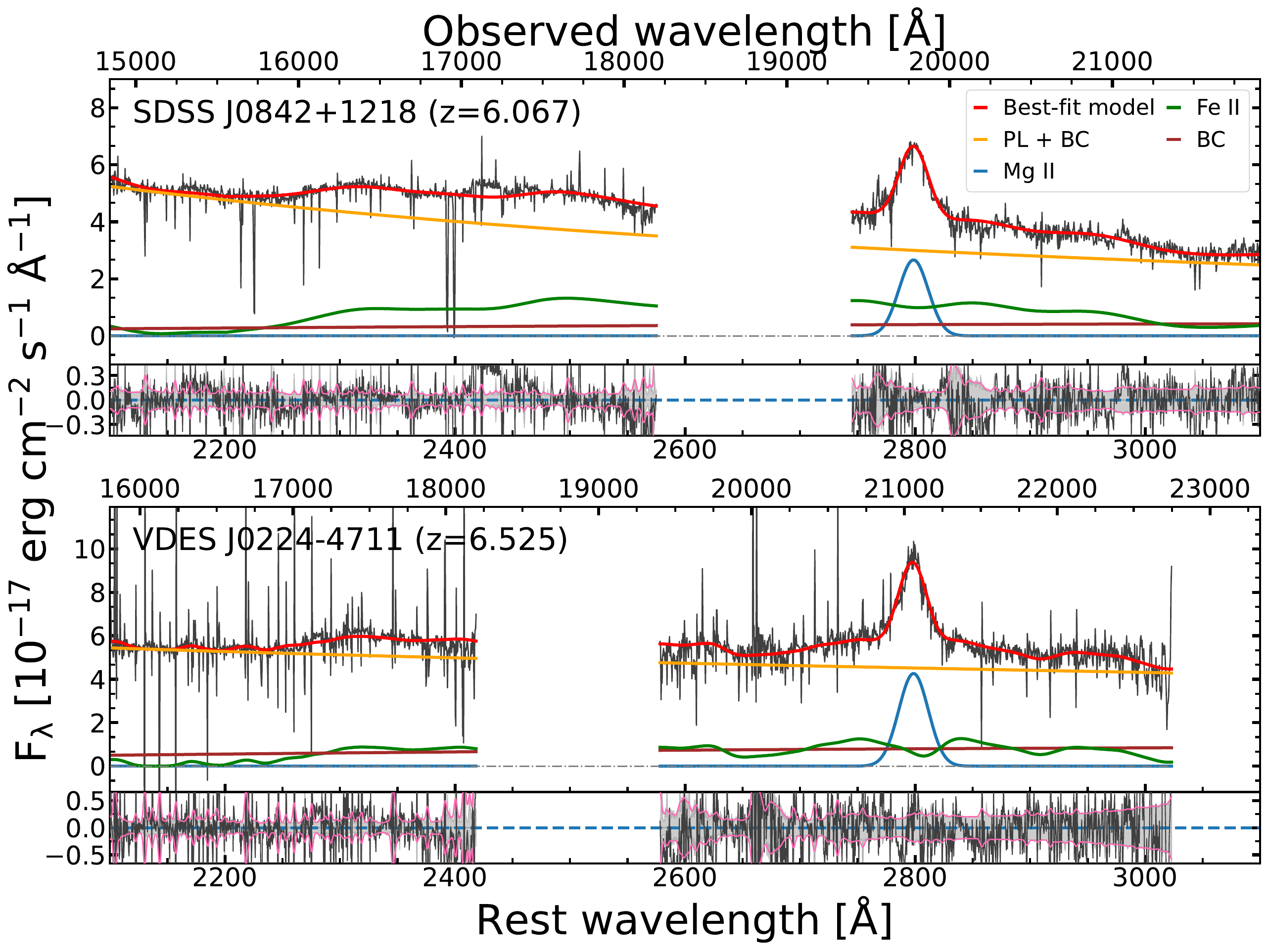}
\caption{Spectral fitting for two representative spectra in the E-XQR-30 sample at $\lambda_{\rm rest}\sim$ $2100 - 3100$ \AA. 
The heavy atmospheric absorption regions at 18,200--19,400 \AA~are masked out in the figure. In each panel, the color-coded lines show the power-law continuum plus Balmer continuum (PL + BC, orange), the Balmer continuum only (BC, brown), the scaled \ion{Fe}{2} template from \cite{2006ApJ...650...57} (\ion{Fe}{2}, green), and the \ion{Mg}{2} emission line (\ion{Mg}{2}, blue). The red line shows the sum of all components (best-fit model). For each spectrum, the residual flux of the best-fit model with the $\pm 1\sigma$ observed errors (grey ranges) is shown in the lower sub-panel. 
We also smooth the errors and use the pink curves to indicate their boundaries.
The full list of the spectra with the best-fit model fits are published as online figures, and the captions of the online figures are the same as the caption of this figure (but for different quasars).
\label{fig:fitted_spec_xqr30}}
\end{figure*}

XQR-30 is an European Southern Observatory (ESO) large program \citep[ID:1103.A-0817, P.I. V. D’Odorico;][]{DOdorico2023} that spent $\sim250$ hours of observing time with the XSHOOTER spectrograph \citep{2011A&A...536A.105V} at the Very Large Telescope (VLT). The program provides high-quality spectra of 30 quasars at $5.8 \leq z \leq 6.6$ in the optical and near-IR wavelengths. These quasars are the brightest $z \geq5.8$ quasars accessible from the Paranal Observatory known by 2018. The mean $J$-band magnitude of this sample is 19.44 mag. The observations were carried out with a $0\farcs9$-wide slit in the VIS (visible) arm and a $0\farcs6$-wide slit in the NIR (near-IR) arm. This configuration yields the nominal spectral resolution of $R\sim8900$ in the optical and $R\sim8100$ in the near-IR.  
An additional set of 12 quasar spectra in the same redshift and luminosity ranges were retrieved from the XSHOOTER archive. These spectra have comparable data quality \citep[e.g.,][]{Becker2015, bosman2018, 2020ApJ...905...51}. 
The combination of this sample and the main XQR-30 sample, named as the \textit{enlarged} XQR-30 sample (hereafter E-XQR-30), consists of 42 spectra with a total exposure times of $\sim350$ hours. The median effective resolving power of E-XQR-30 is $R\sim$ 11,400 and 9800 in the VIS and NIR arms, respectively. The data reduction of the E-XQR-30 sample was performed using a modified version of a custom IDL pipeline \citep{becker2012, 2019ApJ...883...163, 2016A&A...594...A91}. This pipeline provides optimal spectral extraction, telluric correction, and sky subtraction (particularly in the near-IR range), which is optimal for our science goal \citep{noll2012,jones2013}. The E-XQR-30 NIR spectra are not corrected for Galactic extinction, whose effect is typically negligible at the observed wavelength \citep{DOdorico2023}. More details of the E-XQR-30 quasars and data reduction can be found in \citet{DOdorico2023}.

We used the NIR-arm spectra ($0.98$ -- $2.48$ $\micron$) of the E-XQR-30 quasars for our analyses, since the \ion{Fe}{2} and \ion{Mg}{2} emission lines are redshifted to the near-IR wavelengths at $z>6$. These quasars have mean signal-to-noise ratios (SNRs) per pixel above 20 at 2750--2850 \AA.
We removed DELS~J1535+1943 from our final sample because its dust-reddened continuum \citep{Yang21} cannot be modeled with a single power-law function. We also removed SDSS~J0100+2802, one of the external E-XQR-30 quasars discovered by \cite{Wu15}, because its line shape of \ion{Mg}{2} is strongly affected by the atmospheric absorption at 2.05 \micron. There is one gravitationally lensed quasar in our sample, UHS~J0439+1634 \citep{2019ApJL...870...L11}, and we take into account the magnification factor of 51.3 reported in \citet{2019ApJL...870...L11}. 
Based on our visual inspection, we further rejected quasars with strong broad absorption lines and quasars at redshifts $z<6.05$, whose \ion{Mg}{2} emission lines in the spectra are severely affected by the heavy atmospheric absorption at $\sim1.9$ $\micron$.
We also rejected PSO J065-26 and PSO J217-07, because their \ion{Mg}{2} lines are strongly affected by the atmospheric absorption at $\sim2$ $\micron$.
In the end, our E-XQR-30 sample consists of 17 quasars at redshifts between 6.05 and 6.6.

For each selected quasar, its VIS- and NIR-arm spectra were merged with each other using their overlapping spectral region at 1.00 -- 1.02 $\micron$. The merged spectrum was then scaled to match the $J$-band photometry of the quasar. The $J$-band magnitudes of our quasars were collected from the literature \citep[see Table \ref{tab:E-XQR-30};][]{Banados2016,2017ApJ...849...91,2020MNRAS...494...789}.
Finally, we rebinned the spectrum by an integer number of pixels using a public code \textsf{SpectRes} \citep{2017......} so that all quasar spectra have a fixed wavelength step of 50 km s$^{-1}$. This step harmonized the resolution of different spectra to reduce systematic bias.
The bolometric luminosity was derived by applying a bolometric correction of $L_\mathrm{bol}=5.15 \times \lambda L_{3000}$ \citep{shen2008}, where $L_{3000}$ is the luminosity at rest-frame 3000 \AA. The typical error in the bolometric correction is $\sim0.1$ dex \citep{2006ApJS...166...470}, and this error is propagated into the uncertainty estimate for $L_\mathrm{bol}$. The median bolometric luminosity of the final sample of E-XQR-30 is log($L_\mathrm{bol}$[erg s$^{-1}$]) $\sim47.3\pm0.1$, where the uncertainty is the normalized median absolute deviation (NMAD). The measured $L_\mathrm{bol}$ scatter of the E-XQR-30 sample is similar to the propagated uncertainty of each $L_\mathrm{bol}$. The minimum and median values of the mean SNRs per pixel at the rest-frame 2250--2350 \AA~of the rebinned spectra are $40$ and $52$, respectively.
Figure \ref{fig:fitted_spec_xqr30} shows two examples of the final spectra at rest-frame 2100--3100 \AA.

\begin{figure*}[!]
\plotone{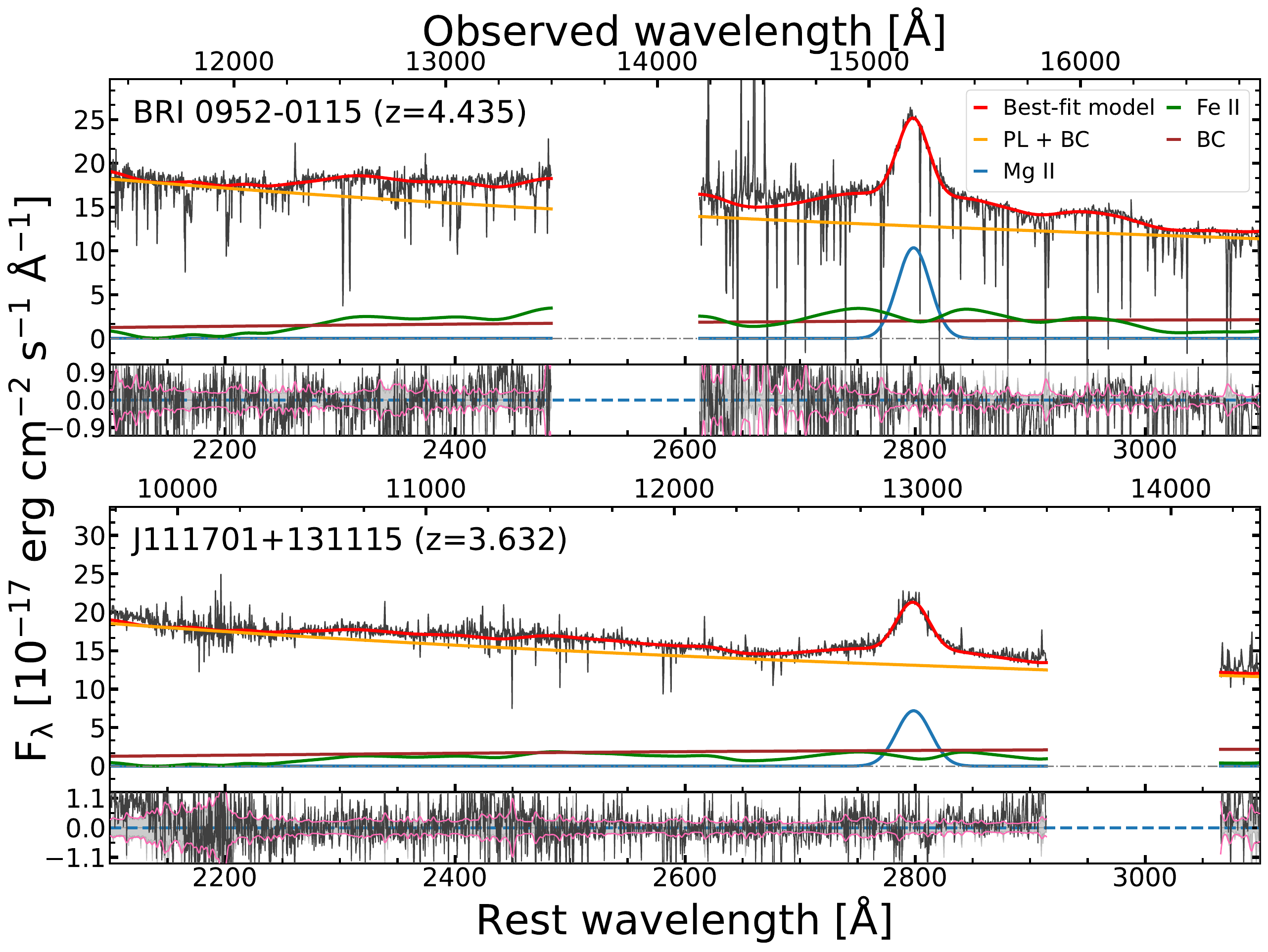}
\caption{Same as Figure~\ref{fig:fitted_spec_xqr30} but for the XQ-100 sample. The heavy atmospheric absorption regions at 13,500--14,200 \AA~are masked out in the figure. The \ion{Fe}{2}/\ion{Mg}{2} ratios of BRI~0952-0115 (upper panel) and J111701+131115 (lower panel) are $5.15_{-0.04}^{+0.04}$ and $3.74_{-0.08}^{+0.09}$, respectively. 
The full list of the spectra with the best model fits are published as online figures. For the XQ-100 spectra that show flux correction problems as described in Section \ref{sec:data_analyses}, they are labeled with Flag=1 in Table \ref{tab:XQ100} and their affected wavelength regions are marked with a red bar in the online figures.
\label{fig:fitted_spec_xq100}}
\end{figure*}

\subsection{The XQ-100 quasar sample} \label{subsec:data_XQ100}

The intermediate-redshift sample is from XQ-100, another VLT/XSHOOTER  large program of ESO (ID:189.A-0424, P.I. S. L\'{o}pez; \citealt{2016A&A...594...A91}). XQ-100 provides 100 high-quality quasar spectra at $z \sim$ 3.5 to 4.8. The adopted slit widths were $1\farcs0$ in the UVB arm and $0\farcs9$ in the VIS and NIR arms, providing the nominal resolution of 5400, 8900, and 5600, respectively. The data reduction of XQ-100 was done using a custom IDL pipeline similar to that for E-XQR-30. The XQ-100 spectra are also not corrected for Galactic extinction for the same reason as E-XQR-30. The reduced data were retrieved from the ESO archive.
From the XQ-100 sample, we selected 78 quasars at $3.5 < z < 4.8$ whose \ion{Mg}{2} emission lines are not severely affected by the atmospheric absorption at 1.35 -- 1.42 $\micron$. We thus removed quasars at $z= 3.82$ -- $4.08$. We also removed another quasar (J0747+2739) whose telluric absorption is not well corrected.  

The XQ-100 spectra were merged, scaled, and rebinned using the same methods as for the E-XQR-30 spectra. 
We referred to the $J$-band photometry reported in existing NIR surveys, including the VISTA Hemisphere Survey \citep{Mcmahon2013}, VISTA Kilo-degree Infrared Galaxy Survey \citep{Edge2013}, UKIRT Hemisphere Survey \citep{Dye2018}, UKIDSS Large Area Survey \citep{Lawrence2007}, and Two Micron All Sky Survey Point Source Catalog \citep{Cutri2003}.
The exact reference of each quasar is reported in Table \ref{tab:XQ100}. The median bolometric luminosities of the XQ-100 sample is log($L_\mathrm{bol}$[erg s$^{-1}$]) $\sim47.5\pm0.2$, which is similar to that of E-XQR-30.
The minimum and median values of the mean SNRs per pixel at rest-frame 2250--2350 \AA~of this sample are $30$ and $70$, respectively, which are comparable to the SNRs of the E-XQR-30 spectra.
Figure \ref{fig:fitted_spec_xq100} shows two examples of the spectra.

\begin{figure*}[!]
\plotone{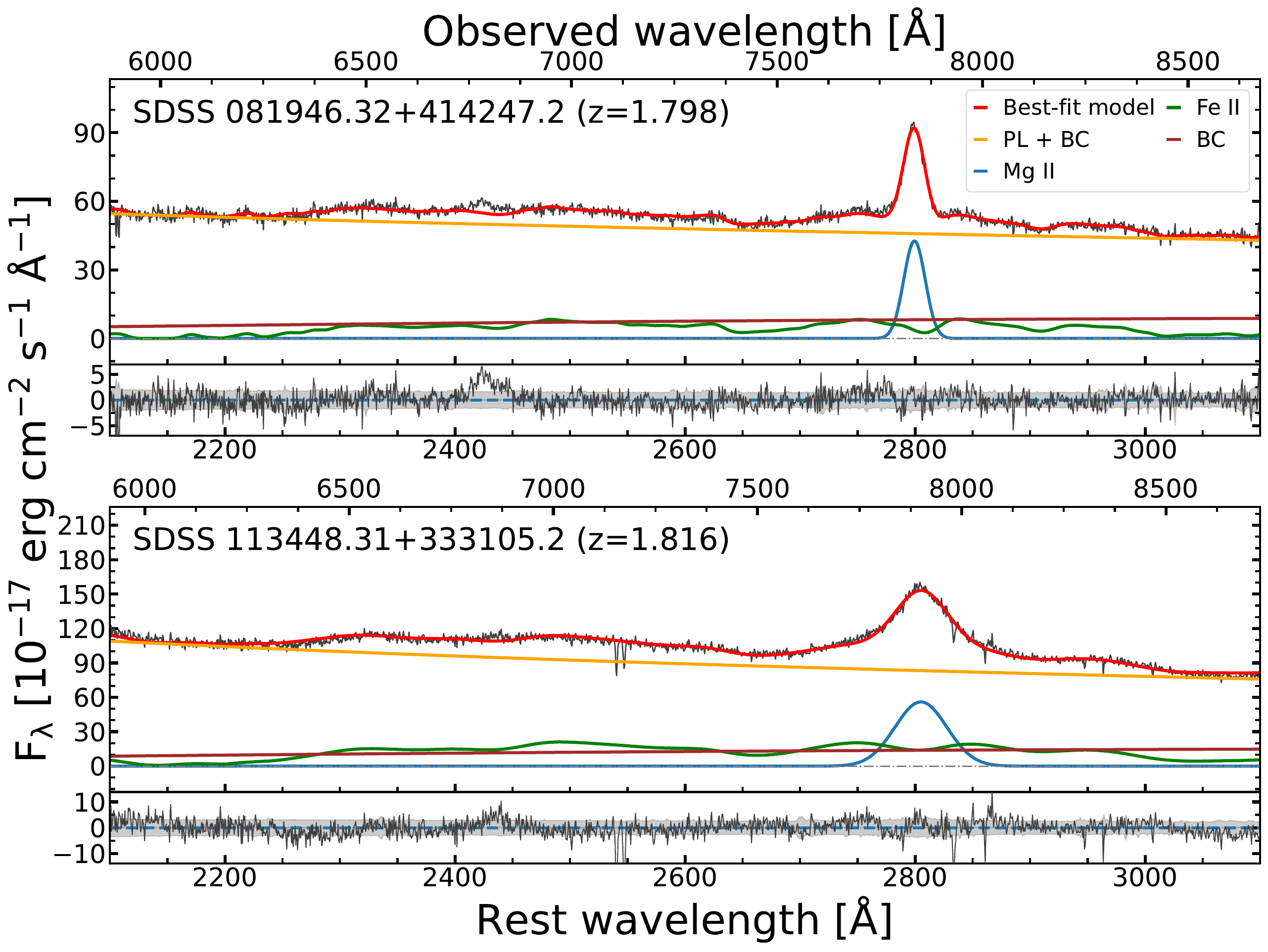}
\caption{Same as Figure~\ref{fig:fitted_spec_xqr30} but for the low-redshift SDSS sample.
The smoothed errors are not shown for SDSS spectra.
The \ion{Fe}{2}/\ion{Mg}{2} ratios of SDSS~081946.32+414247.2 (upper panel) and SDSS~113448.31+333105.2 (lower panel) are $4.31_{-0.10}^{+0.09}$ and $3.75_{-0.07}^{+0.06}$, respectively. 
\label{fig:fitted_spec_sdss}}
\end{figure*}

\subsection{The SDSS DR16 quasar sample} \label{subsec:data_SDSS16Q}

We selected our low-redshift sample from the Data Release 16 of the Sloan Digital Sky Survey quasar catalog  \citep[hereafter SDSS DR16Q;][]{2020ApJS...250...8}. The SDSS DR16Q spectra are all corrected for Galactic extinction using the dust map of \citet{schlegel1998} by \citet{2020ApJS...250...8}. The spectral resolution of the SDSS DR16Q quasar spectra is $R \sim2000$ (150 km s$^{-1}$ per each resolution element), which is lower than the resolutions of E-XQR-30 and XQ-100. We tested the effect of the resolution using $z>6.05$ E-XQR-30 quasar spectra, whose \ion{Mg}{2} emission lines are free from the atmospheric absorption window and thus more closely resemble the SDSS spectral situation. We degraded the E-XQR-30 quasar spectra with a Gaussian kernel to reduce the spectral resolutions from their effective resolutions reported in \cite{DOdorico2023} to 2000. 
The median \ion{Fe}{2}/\ion{Mg}{2} value of the degraded sample is only 2$\%$ lower than that of the original sample, within 0.15 $\sigma$ significance.

The comparison of the \ion{Fe}{2}/\ion{Mg}{2} values is presented in Appendix \ref{sec:degraded E-XQR-30 appendix}. Therefore, we concluded that the relatively low spectral resolution of the SDSS DR16Q sample does not affect our measurements of \ion{Fe}{2}/\ion{Mg}{2}.

We chose SDSS DR16Q quasars that have similar bolometric luminosities as the E-XQR-30 and XQ-100 quasars. We first used the median SNRs of the SDSS spectra provided by the DR16Q catalog to select high-quality spectra, requiring median SNRs higher than 20 per pixel. We also limited our low-redshift range to $0.81 < z < 1.97$ so that the SDSS spectra cover the rest-frame 2100--3100 \AA. This redshift range is slightly different from that of \citet{2017ApJ...834...203}, who  used the SDSS spectra at $0.72 < z < 1.63$ to perform \ion{Fe}{2}/\ion{Mg}{2} measurements. This is because the spectral window for their emission line fitting, the rest-frame 2200--3500 \AA, is different from ours. We will discuss the effect of these two different spectral windows in Section \ref{subsec:systematic uncertainty}. 
For each quasar in the sample, we scaled its spectrum to match the $r$-band photometry of the quasar. 
We then fitted the spectrum to measure $L_{3000}$ and computed the bolometric luminosity from $L_{3000}$.
We selected 1291 SDSS quasars in the bolometric luminosity range of log$(L_\mathrm{bol} \mathrm{[erg\ s^{-1}]}) \sim 47.0$--$48.0$. 

\begin{figure*}[tb]
\plotone{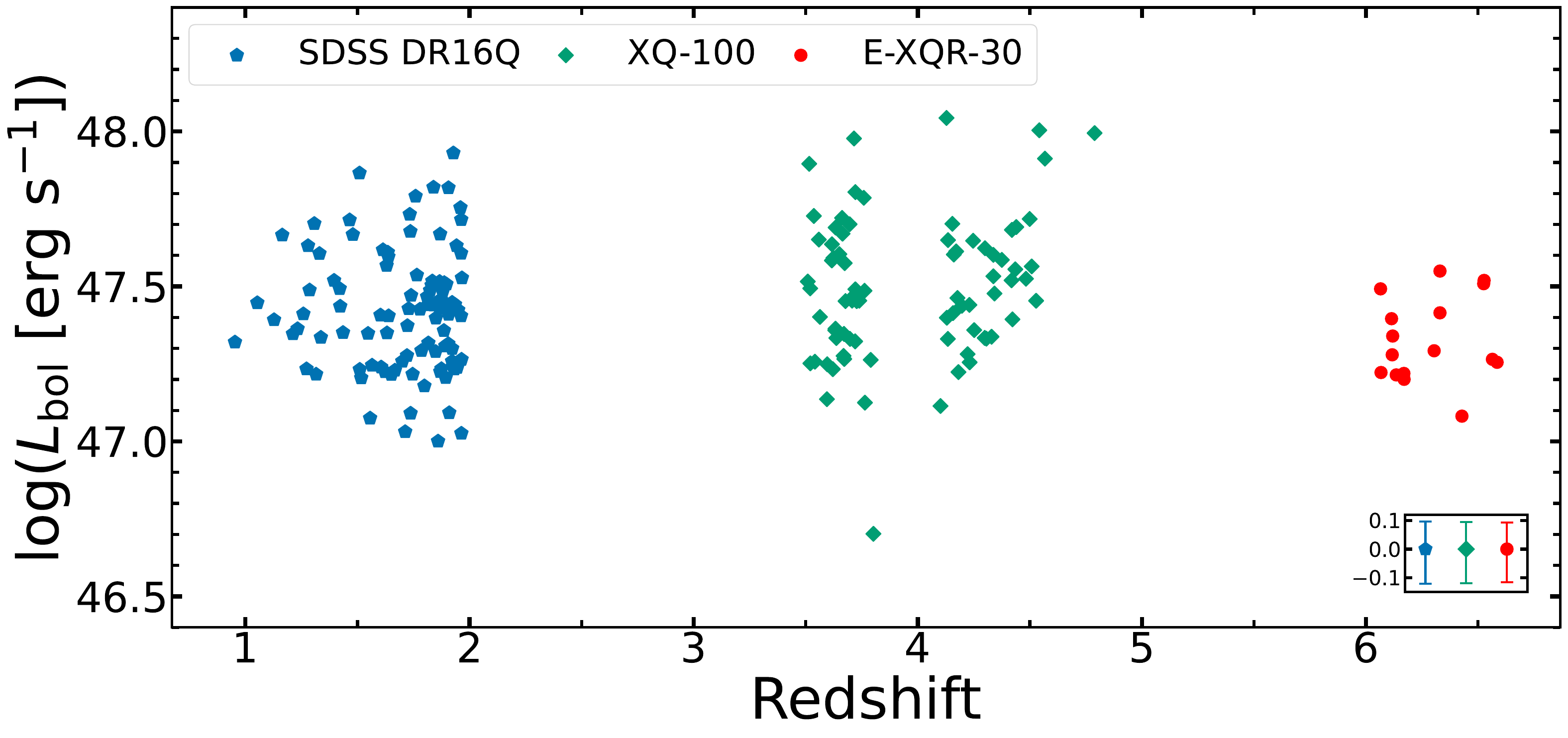}
\caption{The redshift - bolometric luminosity plane of the three quasar samples used in this work. The red circles, green diamonds, and blue pentagons represent the E-XQR-30 quasars at $6.0\leq z\leq6.6$, the XQ-100 quasars at $3.5\leq z\leq 4.8$, and the low-redshift SDSS  quasars, respectively. Their typical measurement errors of $\log{L_\mathrm{bol}}$ are shown in the lower right corner.
These quasars are among the most luminous quasar populations at these epochs ($47.0\la \log{L_\mathrm{bol}}\ \mathrm{[erg\ s^{-1}]} \la48.0$).}
\label{fig:L_bol_Z}
\end{figure*}

We selected our final low-redshift sample from these 1291 high-luminosity SDSS quasars, forcing it to follow the luminosity distribution of the E-XQR-30 and XQ-100 samples. This procedure enables us to have a better control over the three quasar samples at different redshift ranges.
To achieve this goal, we extracted a luminosity-matched sample of 99 quasars from the SDSS quasars. The selected sample size is limited by the small number of high-luminosity quasars (log$(L_\mathrm{bol} \mathrm{[erg\ s^{-1}]})>47.5$) in the 1291 SDSS quasars.
These 99 SDSS quasars span a redshift range of $0.95 \leq z \leq 1.97$, and their minimum and median values of the mean SNRs per pixel at rest-frame 2250--2350 \AA\ are 24 and 39, respectively. Figure \ref{fig:fitted_spec_sdss} shows two examples of the spectra. Figure \ref{fig:L_bol_Z} shows the bolometric luminosity distribution of the three samples as a function of redshift.

\section{Spectral Analyses} \label{sec:data_analyses}

We carried out multi-component spectral fitting for the three quasar samples introduced in Section \ref{sec:observation}. For each quasar, its observed continuum was decomposed into a power-law continuum ($F_{\rm PL}$= $F_0 \lambda^{\alpha_{\lambda}}$), a Balmer continuum ($F_{\rm BC}$), and an iron pseudo-continuum, with three free parameters, namely the scale factor and the slope of a single power-law function, and the scale factor of the UV iron template.
To measure the \ion{Fe}{2} flux, we used the empirical iron template from \citet[hereafter the T06 iron template]{2006ApJ...650...57}. The T06 iron template is based on a high resolution spectrum of I Zw I, a narrow-line Seyfert 1 galaxy at $z=0.061$. This template has been frequently used in the literature \citep[e.g.,][]{2017ApJ...834...203, 2019ApJ...874...22, 2021ApJ...917...107, 2020ApJ...905...51, 2020ApJ...898...105, 2022ApJ...925...121}. It is also suited for our analyses, because it takes into account the \ion{Fe}{2} contribution underneath \ion{Mg}{2} based on photoionization models. The iron template was convolved with Gaussian kernels to generate a range of line widths of FWHM $=$ 500--10,000 km s$^{-1}$, with a step size of 500 km s$^{-1}$.

For the underlying Balmer continuum, we adopted the following formula introduced by \citet{1982ApJ....255...25},
\begin{equation}
F_{\rm BC}=F_{\rm BE} \, B_{\lambda}(T_{e}) \, (1-e^{-\tau_{e}(\lambda/\lambda_{\rm BE})^3}),
\end{equation}
where $B_{\lambda}(T_{e})$ is the Planck function at an electron temperature $T_{e}$ and $\tau_{e}$ is the optical depth at the Balmer edge $\lambda_{\rm BE}=3646$ \AA. 
Owing to the limited wavelength coverage, we followed previous studies \citep[e.g.,][]{2003ApJ....596...817, 2007ApJ...669...32, 2011ApJ...739...56, 2017ApJ...849...91, 2019ApJ...874...22, 2020ApJ...905...51, 2020ApJ...898...105, 2022ApJ...925...121} and fixed the normalization factor $F_{\rm BE}$ to be $30\%$ of $F_{\rm PL}( \lambda=3675$ \AA).
The two parameters $T_{e}$ and $\tau_{e}$ were fixed to be 15,000 K and 1, respectively, following the literature \citep[e.g.,][]{2011ApJ...739...56,2017ApJ...849...91, 2019ApJ...874...22, 2020ApJ...905...51, 2020ApJ...898...105}. 
These choices can help us avoid systematic differences when we compare with previous results later.
 
We generated a model to fit the spectrum with the single power-law continuum, the Balmer continuum, and the T06 iron template. The fitting method is similar to that described in \cite{2020ApJ...898...105}, and is slightly modified to suit our three samples and science goals. The VIS and NIR arms of XSHOOTER cover a wide wavelength range of 550--2480 nm. To focus on \ion{Fe}{2} and \ion{Mg}{2} in this study and optimize our results, we chose to fit the NIR-arm spectrum only. The employed continuum$+$iron fitting windows are $\lambda_{\rm rest}=2100-2750$ \AA~ and $2850-3100$ \AA. The robustness of the continuum measurement in this spectral range is discussed in Section \ref{subsec:continuum robustness}. We excluded the strong atmospheric absorption ranges for each E-XQR-30 and XQ-100 spectrum, and modified the longest wavelengths fitted for some spectra due to their limited spectral coverage. 
A small number of the XQ-100 spectra displayed a glitch in the connection region between the NIR and VIS arms, due to a wavelength-dependent flux correction problem. This will likely affect our measurements of the continuum and \ion{Fe}{2}/\ion{Mg}{2} values. We visually checked the spectra and adjusted the continuum$+$iron fitting windows for these spectra. Wavelength regions with suspicious arm-connection problems were excluded. We also flagged them in the XQ-100 online table (see Table \ref{tab:XQ100}). All best-fit models performed well near \ion{Mg}{2} and in most of the \ion{Fe}{2} measurement regions. The results change by only $\sim 1\%$ without these flagged spectra, suggesting that this has very limited impact on our study. For completeness, we still used the full XQ-100 sample in our analyses.
The low-redshift SDSS sample did not change because it has no strong atmospheric absorption bands. 
These fitting windows cover some weak emission lines such as \ion{C}{2}]~$\lambda$2326 and [\ion{O}{2}]~$\lambda$2471, but the distribution of the iron pseudo-continuum is much wider than these small emission lines, and thus they do not affect our spectra fitting results.

The quasar spectra were iteratively fitted to obtain the power-law scale and slope factors, and the scale factor of the iron template. We redshifted the iron templates to the \ion{Mg}{2}-based redshift, without considering a potential velocity shift. The fitting steps on the continuum$+$iron windows are described below.
\begin{enumerate}
\item A single power-law continuum plus Balmer continuum were fitted to the spectrum. 
\item The iron template was fitted to the continuum-subtracted spectrum.
\item The continuum model was updated by repeating the continuum fitting to the spectrum after subtracting the iron model.
\end{enumerate}
All fitting steps above used the least-square method. The second and third steps were repeated until the three parameters achieved $< 1\%$ convergence. In the procedure, we used all iron templates with different line widths, and selected one iron model that returned the minimum $\chi^2$ value at the continuum$+$iron windows. The broad \ion{Mg}{2} emission line was then fitted to the residual spectrum after subtracting the best-fit continuum$+$iron model. A single Gaussian profile was used to fit the emission line. The two \ion{Mg}{2} double emission lines mix together due to the broad line width, so a single Gaussian profile is sufficient for the fit. The fitting window for \ion{Mg}{2} was $\lambda_{\rm rest}=2750$--$2850$ \AA. 

We ran the above procedure twice. After the completion of the first iteration, we used the best-fit models to perform a single round of 3 sigma-clipping to mask out the pixels that could be significantly affected by strong sky lines or cosmic rays. The sigma is the observed flux error. We then performed the second iteration after the sigma clipping. Figures \ref{fig:fitted_spec_xqr30},  \ref{fig:fitted_spec_xq100}, and \ref{fig:fitted_spec_sdss} show a few examples of the high-redshift E-XQR-30, intermediate-redshift XQ-100, and low-redshift SDSS spectra with their best-fit models, respectively.

The \ion{Fe}{2} flux was calculated by integrating the scaled iron template over 2200  $\leq$ $\lambda_{\rm rest}$ [\AA] $\leq$ 3090. The equivalent widths (EWs) of the \ion{Fe}{2} and \ion{Mg}{2} emission lines were derived by dividing the line flux by the power-law plus Balmer continuum flux at $\lambda_{\rm rest}=$ 3000 \AA. This procedure is to make a direct comparison with \cite{2017ApJ...834...203} and \cite{2020ApJ...898...105}, who measured EWs in the same way. 
We also calculated the \ion{Fe}{2} and \ion{Mg}{2} EWs using the standard EW measurement. The differences of the measured EWs from the two methods are only 0.06--0.08 dex, and are much smaller than the measurement errors.
The monochromatic luminosity $\lambda L_{3000}$ and bolometric luminosity $L_\mathrm{bol}$ were measured using the method described  in Section \ref{subsec:data_XQR30}.

We propagated the measurement uncertainties using Monte Carlo simulations. 
Specifically, we generated 1000 mock spectra for each quasar. The flux at each wavelength was the original flux plus an error flux. The error flux was randomly drawn from a Gaussian distribution with $\sigma$ equal to the observed error at this wavelength.
We repeated our fitting process for each version of the noise-added spectrum. We obtained a distribution of the 1000 results for each continuum+emission parameter, and their $1\sigma$ uncertainties were measured from the 16th and 84th percentiles of the distribution.

\section{Results} \label{sec:results}

\subsection{Emission Line Properties} \label{subsec:properties}

The \ion{Fe}{2} and \ion{Mg}{2} broad emission line properties are derived from our best-fit models. With our \ion{Mg}{2} measurement using the T06 iron template, we estimate the Eddington luminosities of the quasars in all samples by the following formula,
\begin{equation}
L_{\text{Edd}} = 1.3 \times 10^{38} \left(\frac{M_{\text{BH}}}{M_{\odot}}\right)\mathrm{erg\ s^{-1}},
\end{equation}
where $M_\mathrm{BH}$ is the BH mass calculated using the virial relation from \cite{2009ApJ...699...800}. 
The Eddington ratio $L_\mathrm{bol}/L_\mathrm{Edd}$ is obtained by dividing the bolometric luminosity by the Eddington luminosity. 
The uniform measurements of the Eddington ratios for our three redshift samples help us correct the correlation between the \ion{Fe}{2}/\ion{Mg}{2} ratio and the Eddington ratio (see Section \ref{subsec:EddRatio_correlation}). 
The measurement uncertainties of these quantities are derived from the same Monte Carlo simulations as described in Section \ref{sec:data_analyses}. We calculate these ratios from the 1000 mock spectra of each quasar and obtain their distributions. The $1\sigma$ uncertainties are measured from the 16th and 84th percentiles of the distributions.

The measured properties of the E-XQR-30 quasars are presented in Table \ref{tab:E-XQR-30}. 
The emission line properties of the low-redshift and intermediate-redshift samples are measured in the same way. The measured properties of the XQ-100 and SDSS samples are shown in Table \ref{tab:XQ100} and Table \ref{tab:SDSS} respectively, with machine-readable formats available online.

\movetabledown=45mm
\begin{rotatetable}
\begin{deluxetable*}{lccccccccccc}
\tablecaption{The measured properties of the E-XQR-30 quasars\label{tab:E-XQR-30}}
\tablewidth{0pt}
\tablehead{
\colhead{ID} & \colhead{$z_\mathrm{MgII}$} & \colhead{$J_\mathrm{AB}$}& \colhead{Ref} & \colhead{EW \ion{Mg}{2}} & \colhead{FWHM \ion{Mg}{2}} & \colhead{EW \ion{Fe}{2}} & \colhead{\ion{Fe}{2}/\ion{Mg}{2}} & \colhead{log($\lambda L_{3000}$)} & \colhead{log($L_\mathrm{bol}$)} & \colhead{$L_\mathrm{bol}/L_\mathrm{Edd}$} & \colhead{(\ion{Fe}{2}/\ion{Mg}{2})$_c$}\\
\colhead{} & \colhead{} & \colhead{} & \colhead{} & \colhead{[\AA]} & \colhead{[km s$^{-1}$]} & \colhead{[\AA]} & \colhead{} & \colhead{[erg s$^{-1}$]} & \colhead{[erg s$^{-1}$]} & \colhead{} & \colhead{}
}
\decimalcolnumbers
\startdata
PSO J158-14 & 6.065 & 19.27 & 3 & $19.31^{+0.22}_{-0.23}$ & $2711^{+30}_{-30}$ & $143.0^{+2.3}_{-2.1}$ & $7.41^{+0.13}_{-0.12}$ & $46.780^{+0.001}_{-0.002}$ & $47.49^{+0.09}_{-0.11}$ & $1.83^{+0.45}_{-0.42}$ & $4.55^{+0.33}_{-0.26}$ \\
SDSS J0842+1218 & 6.067 & 19.70 & 3 & $32.46^{+0.30}_{-0.28}$ & $3238^{+28}_{-27}$ & $290.4^{+3.0}_{-3.4}$ & $8.95^{+0.11}_{-0.10}$ & $46.511^{+0.002}_{-0.002}$ & $47.22^{+0.09}_{-0.12}$ & $0.94^{+0.22}_{-0.22}$ & $6.45^{+0.42}_{-0.35}$ \\
PSO J239-07 & 6.114 & 19.35 & 1 & $29.37^{+0.28}_{-0.25}$ & $3386^{+30}_{-33}$ & $244.6^{+2.8}_{-2.4}$ & $8.33^{+0.09}_{-0.09}$ & $46.684^{+0.001}_{-0.002}$ & $47.40^{+0.10}_{-0.12}$ & $1.05^{+0.27}_{-0.24}$ & $5.84^{+0.42}_{-0.31}$ \\
PSO J217-16 & 6.135 & 19.69 & 1 & $19.36^{+0.30}_{-0.26}$ & $3212^{+42}_{-46}$ & $182.4^{+3.1}_{-2.9}$ & $9.42^{+0.18}_{-0.17}$ & $46.503^{+0.002}_{-0.002}$ & $47.21^{+0.10}_{-0.12}$ & $0.95^{+0.25}_{-0.23}$ & $6.78^{+0.50}_{-0.39}$ \\
PSO J359-06 & 6.169 & 19.85 & 1 & $21.43^{+0.36}_{-0.31}$ & $2446^{+38}_{-35}$ & $287.0^{+3.2}_{-3.5}$ & $13.39^{+0.24}_{-0.23}$ & $46.508^{+0.002}_{-0.002}$ & $47.22^{+0.10}_{-0.11}$ & $1.64^{+0.42}_{-0.36}$ & $8.44^{+0.54}_{-0.49}$ \\
PSO J060+24 & 6.170 & 19.95 & 3 & $17.86^{+0.31}_{-0.34}$ & $2683^{+53}_{-54}$ & $141.6^{+2.9}_{-3.1}$ & $7.93^{+0.19}_{-0.20}$ & $46.489^{+0.002}_{-0.002}$ & $47.20^{+0.09}_{-0.11}$ & $1.33^{+0.33}_{-0.30}$ & $5.25^{+0.37}_{-0.32}$ \\
VDES J2211-3206 & 6.330 & 19.54 & 3 & $28.86^{+0.41}_{-0.41}$ & $3123^{+41}_{-40}$ & $229.1^{+5.3}_{-5.3}$ & $7.94^{+0.14}_{-0.17}$ & $46.703^{+0.003}_{-0.003}$ & $47.41^{+0.10}_{-0.11}$ & $1.26^{+0.31}_{-0.28}$ & $5.33^{+0.34}_{-0.32}$ \\
PSO J183+05 & 6.428 & 19.77 & 2 & $31.70^{+0.69}_{-0.67}$ & $3502^{+76}_{-79}$ & $326.1^{+7.1}_{-7.4}$ & $10.29^{+0.30}_{-0.28}$ & $46.370^{+0.003}_{-0.003}$ & $47.08^{+0.10}_{-0.13}$ & $0.68^{+0.19}_{-0.16}$ & $8.00^{+0.61}_{-0.48}$ \\
VDES J0224-4711 & 6.525 & 19.73 & 3 & $31.66^{+0.36}_{-0.33}$ & $3239^{+29}_{-30}$ & $152.2^{+3.2}_{-3.2}$ & $4.81^{+0.09}_{-0.09}$ & $46.797^{+0.002}_{-0.002}$ & $47.51^{+0.10}_{-0.10}$ & $1.30^{+0.34}_{-0.28}$ & $3.20^{+0.21}_{-0.19}$ \\
PSO J231-20 & 6.564 & 19.60 & 3 & $23.06^{+0.46}_{-0.50}$ & $4179^{+81}_{-68}$ & $278.6^{+8.5}_{-6.7}$ & $12.08^{+0.37}_{-0.27}$ & $46.553^{+0.003}_{-0.004}$ & $47.26^{+0.09}_{-0.13}$ & $0.59^{+0.13}_{-0.16}$ & $9.73^{+0.76}_{-0.49}$ \\
PSO J323+12 & 6.585 & 19.63 & 3 & $37.81^{+0.68}_{-0.60}$ & $2764^{+41}_{-41}$ & $303.5^{+8.0}_{-7.0}$ & $8.03^{+0.18}_{-0.15}$ & $46.543^{+0.003}_{-0.004}$ & $47.26^{+0.09}_{-0.12}$ & $1.34^{+0.30}_{-0.33}$ & $5.31^{+0.37}_{-0.28}$ \\
\cline{0-11}
ULAS J1319+0950 & 6.117 & 19.58 & 3 & $21.34^{+0.37}_{-0.32}$ & $3989^{+68}_{-67}$ & $114.2^{+2.6}_{-3.0}$ & $5.35^{+0.15}_{-0.14}$ & $46.568^{+0.002}_{-0.002}$ & $47.28^{+0.09}_{-0.14}$ & $0.66^{+0.15}_{-0.18}$ & $4.20^{+0.32}_{-0.21}$ \\
CFHQS J1509-1749 & 6.119 & 19.75 & 3 & $16.88^{+0.24}_{-0.23}$ & $3448^{+50}_{-53}$ & $42.1^{+2.0}_{-2.1}$ & $2.49^{+0.13}_{-0.11}$ & $46.628^{+0.002}_{-0.002}$ & $47.34^{+0.09}_{-0.13}$ & $0.95^{+0.22}_{-0.26}$ & $1.79^{+0.17}_{-0.11}$ \\
SDSS J1030+0524 & 6.304 & 19.87 & 3 & $26.99^{+0.36}_{-0.29}$ & $3161^{+38}_{-36}$ & $141.3^{+3.4}_{-3.0}$ & $5.24^{+0.12}_{-0.11}$ & $46.581^{+0.002}_{-0.002}$ & $47.29^{+0.09}_{-0.12}$ & $1.07^{+0.26}_{-0.25}$ & $3.66^{+0.26}_{-0.21}$ \\
VST-ATLAS J025-33 & 6.330 & 19.10 & 3 & $18.90^{+0.22}_{-0.23}$ & $3471^{+40}_{-42}$ & $74.0^{+2.2}_{-2.1}$ & $3.91^{+0.10}_{-0.10}$ & $46.838^{+0.002}_{-0.001}$ & $47.55^{+0.09}_{-0.11}$ & $1.19^{+0.26}_{-0.28}$ & $2.66^{+0.22}_{-0.14}$ \\
WISEA J0439+1634 & 6.520 & 17.47 & 3 & $17.60^{+0.06}_{-0.05}$ & $2982^{+11}_{-11}$ & $122.3^{+0.3}_{-0.4}$ & $6.95^{+0.03}_{-0.03}$ & $45.884^{+0.000}_{-0.000}$ & $46.60^{+0.09}_{-0.12}$ & $0.54^{+0.13}_{-0.13}$ & $5.73^{+0.36}_{-0.30}$ \\
PSO J036+03 & 6.527 & 19.40 & 3 & $5.43^{+0.26}_{-0.26}$ & $2556^{+112}_{-104}$ & $44.0^{+3.7}_{-3.5}$ & $8.11^{+0.65}_{-0.61}$ & $46.808^{+0.003}_{-0.003}$ & $47.52^{+0.10}_{-0.11}$ & $2.12^{+0.62}_{-0.50}$ & $4.81^{+0.58}_{-0.42}$ \\
\enddata
\tablerefs{1 - \cite{Banados2016}; 2 - \cite{2017ApJ...849...91}; 3 - \cite{2020MNRAS...494...789}.}
\tablecomments{
The \ion{Mg}{2} redshifts of the quasars are given in column (2). The $J$-band magnitudes and their references are listed in cols. (3) and (4). The measured FWHM \ion{Mg}{2} in unit of km s$^{-1}$ and EW \ion{Mg}{2} in unit of \AA~are listed in cols. (5) and (6). The measured EW \ion{Fe}{2} in unit of \AA~is given in col. (7). 
Since the EWs in this study were derived from scaled flux, the \ion{Fe}{2}/\ion{Mg}{2} (col. [9]) are equal to the flux ratios.
The rest-frame continuum luminosity $\lambda L_{3000}$, the bolometric luminosity $L_\mathrm{bol}$ and the Eddington ratio $L_\mathrm{bol}/L_\mathrm{Edd}$ are listed in cols. (9) to (11). 
The \ion{Fe}{2}/\ion{Mg}{2} results after correcting the Eddington ratio dependence are listed in col. (12).
The measurement uncertainties were derived from a Monte Carlo simulation based on the flux error at each wavelength step.
}
\end{deluxetable*}
\end{rotatetable}
\clearpage

\movetabledown=45mm
\begin{rotatetable}
\begin{deluxetable*}{lcccccccccccc}
\tabletypesize{\footnotesize} 
\tablecaption{The measured properties of the XQ-100 quasars \label{tab:XQ100}}
\tablewidth{0pt}
\tablehead{
\colhead{ID} & \colhead{$z_\mathrm{MgII}$} & \colhead{$J_\mathrm{AB}$}& \colhead{Ref} & \colhead{Flag} & \colhead{EW \ion{Mg}{2}} & \colhead{FWHM \ion{Mg}{2}} & \colhead{EW \ion{Fe}{2}} & \colhead{\ion{Fe}{2}/\ion{Mg}{2}} & \colhead{log($\lambda L_{3000}$)} & \colhead{log($L_\mathrm{bol}$)} & \colhead{$L_\mathrm{bol}/L_\mathrm{Edd}$} & \colhead{(\ion{Fe}{2}/\ion{Mg}{2})$_c$} \\
\colhead{} & \colhead{} & \colhead{[mag]} & \colhead{} & \colhead{} & \colhead{[\AA]} & \colhead{[km s$^{-1}$]} & \colhead{[\AA]} & \colhead{} & \colhead{[erg s$^{-1}$]} & \colhead{[erg s$^{-1}$]} & \colhead{} & \colhead{} }
\decimalcolnumbers
\startdata
J133254+005250 & 3.509 & 17.84 & 2 & 1 & $18.61^{+0.09}_{-0.08}$ & $2511^{+13}_{-15}$ & $89.6^{+1.0}_{-1.5}$ & $4.81^{+0.06}_{-0.07}$ & $46.804^{+0.001}_{-0.001}$ & $47.52^{+0.09}_{-0.13}$ & $2.19^{+0.50}_{-0.54}$ & $2.83^{+0.20}_{-0.16}$ \\
HB89 1159+123 & 3.516 & 16.71 & 2 & 0 & $30.48^{+0.09}_{-0.09}$ & $3119^{+10}_{-10}$ & $254.9^{+1.3}_{-1.3}$ & $8.36^{+0.04}_{-0.04}$ & $47.184^{+0.001}_{-0.001}$ & $47.90^{+0.10}_{-0.11}$ & $2.20^{+0.56}_{-0.50}$ & $4.91^{+0.35}_{-0.27}$ \\
SDSS J1024+1819 & 3.520 & 17.88 & 5 & 0 & $21.35^{+0.23}_{-0.27}$ & $4048^{+43}_{-53}$ & $95.1^{+3.0}_{-3.2}$ & $4.45^{+0.15}_{-0.13}$ & $46.782^{+0.001}_{-0.002}$ & $47.49^{+0.09}_{-0.12}$ & $0.82^{+0.20}_{-0.20}$ & $3.32^{+0.28}_{-0.19}$ \\
J101818+054822 & 3.521 & 18.50 & 2 & 1 & $27.04^{+0.16}_{-0.15}$ & $3575^{+23}_{-21}$ & $109.0^{+2.0}_{-2.0}$ & $4.03^{+0.07}_{-0.07}$ & $46.540^{+0.001}_{-0.001}$ & $47.25^{+0.09}_{-0.13}$ & $0.80^{+0.17}_{-0.21}$ & $3.02^{+0.22}_{-0.16}$ \\
J144250+092001 & 3.536 & 17.38 & 2 & 1 & $19.44^{+0.10}_{-0.11}$ & $5118^{+34}_{-33}$ & $60.1^{+1.1}_{-1.3}$ & $3.09^{+0.06}_{-0.06}$ & $47.015^{+0.001}_{-0.001}$ & $47.73^{+0.10}_{-0.11}$ & $0.67^{+0.17}_{-0.15}$ & $2.41^{+0.16}_{-0.13}$ \\
\enddata
\tablerefs{1 - \cite{Cutri2003}, 2 - \cite{Lawrence2007}, 3 - \cite{Edge2013}, 4 - \cite{Mcmahon2013}, 5 - \cite{Dye2018}.}
\tablecomments{The same table as Table \ref{tab:E-XQR-30} but for the XQ-100 sample. The col. (5) lists the flags of whether there is a suspicious arm connection problem in this spectrum at $<2200$ \AA, with 0 indicating normal and 1 indicating a problem. The full table is published in a machine-readable format and only a fraction of the XQ-100 sample are presented here.}
\end{deluxetable*}
\end{rotatetable}

\movetabledown=45mm
\begin{rotatetable}
\begin{deluxetable*}{lcccccccccc}
\tabletypesize{\footnotesize} 
\tablecaption{The measured properties of the SDSS quasars \label{tab:SDSS}}
\tablewidth{0pt}
\tablehead{
\colhead{ID} & \colhead{$z_\mathrm{MgII}$} & \colhead{r$^a$} & \colhead{EW \ion{Mg}{2}} & \colhead{FWHM \ion{Mg}{2}} & \colhead{EW \ion{Fe}{2}} & \colhead{\ion{Fe}{2}/\ion{Mg}{2}} & \colhead{log($\lambda L_{3000}$)} & \colhead{log($L_\mathrm{bol}$)} & \colhead{$L_\mathrm{bol}/L_\mathrm{Edd}$} & \colhead{(\ion{Fe}{2}/\ion{Mg}{2})$_c$} \\
\colhead{} & \colhead{} & \colhead{[mag]} & \colhead{[\AA]} & \colhead{[km s$^{-1}$]} & \colhead{[\AA]} & \colhead{} & \colhead{[erg s$^{-1}$]} & \colhead{[erg s$^{-1}$]} & \colhead{} & \colhead{}
}
\decimalcolnumbers
\startdata
SDSS135823.98+021343.8 & 0.953 & 15.73 & $12.49^{+0.14}_{-0.17}$ & $4374^{+76}_{-77}$ & $82.1^{+1.4}_{-1.3}$ & $6.57^{+0.15}_{-0.13}$ & $46.610^{+0.000}_{-0.000}$ & $47.32^{+0.11}_{-0.10}$ & $0.58^{+0.16}_{-0.12}$ & $5.32^{+0.34}_{-0.32}$ \\
SDSS142107.56+253821.0 & 1.053 & 15.58 & $18.97^{+0.15}_{-0.15}$ & $2666^{+24}_{-31}$ & $109.7^{+1.5}_{-1.6}$ & $5.78^{+0.09}_{-0.09}$ & $46.740^{+0.000}_{-0.000}$ & $47.45^{+0.09}_{-0.13}$ & $1.79^{+0.44}_{-0.46}$ & $3.57^{+0.28}_{-0.21}$ \\
SDSS082045.38+130618.9 & 1.128 & 15.92 & $13.23^{+0.12}_{-0.11}$ & $2810^{+25}_{-28}$ & $86.0^{+1.4}_{-1.5}$ & $6.50^{+0.12}_{-0.12}$ & $46.680^{+0.000}_{-0.000}$ & $47.39^{+0.10}_{-0.13}$ & $1.52^{+0.38}_{-0.39}$ & $4.18^{+0.31}_{-0.25}$ \\
SDSS120858.01+454035.4 & 1.164 & 15.28 & $19.15^{+0.16}_{-0.15}$ & $4256^{+46}_{-42}$ & $96.7^{+1.5}_{-1.5}$ & $5.05^{+0.09}_{-0.09}$ & $46.950^{+0.000}_{-0.000}$ & $47.67^{+0.09}_{-0.12}$ & $0.91^{+0.20}_{-0.22}$ & $3.67^{+0.24}_{-0.21}$ \\
SDSS143120.53+395241.5 & 1.212 & 16.19 & $21.63^{+0.22}_{-0.24}$ & $5563^{+68}_{-63}$ & $95.2^{+2.2}_{-1.9}$ & $4.40^{+0.11}_{-0.10}$ & $46.640^{+0.000}_{-0.000}$ & $47.35^{+0.10}_{-0.12}$ & $0.37^{+0.09}_{-0.09}$ & $3.97^{+0.29}_{-0.23}$ \\
\enddata
\tablenotetext{a}{The $r$-band magnitudes are quoted from \cite{aihara2011}.}
\tablecomments{The same table as Table \ref{tab:E-XQR-30} but for the low-redshift SDSS sample.
The full table is published in the machine-readable format.}
\end{deluxetable*}
\end{rotatetable}
\clearpage

\subsection{Dependence on the Eddington Ratio} \label{subsec:EddRatio_correlation}

The \ion{Fe}{2}/\ion{Mg}{2} flux ratio has been used as a proxy for the Fe/Mg abundance ratio (i.e., [Fe/$\alpha$]). It is expected to show a dramatic change at cosmic age $\sim$ 1 Gyr due to the different evolution timescales of SNe Ia and SNe II, the major producers of Fe and Mg respectively \citep[e.g.,][]{1993ApJ...418...11H, 1999ARAA...37...487}. On the other hand, observations have shown that the \ion{Mg}{2} EW anti-correlates with the Eddington ratio  \citep{2011ApJ...736...86, 2017ApJ...834...203}.
This dependence on the non-abundance parameter has to be corrected when one addresses the chemical enrichment of the BLR cloud.
Photoionization models show that this Eddington ratio dependence can be attributed to the different origin of the \ion{Fe}{2} and \ion{Mg}{2} emission.
\citet{2017ApJ...834...203} argued that \ion{Mg}{2} mainly originates from the fully ionized zone (FIZ), while \ion{Fe}{2} mainly originates from the partially ionized zone (PIZ). When the gas density increases, the electron temperature of the FIZ decreases, the fraction of Mg$^{+}$ increases, and thus the \ion{Mg}{2} emitted by the collisional excitation is stronger. Meanwhile, the electron temperature of the PIZ and its Fe$^{+}$ and Mg$^{+}$ fractions are basically unchanged.
In order to correct for this density effect, \citet{2017ApJ...834...203} introduced a correction to take into account the Eddington ratio dependence of the observed \ion{Fe}{2}/\ion{Mg}{2} ratio:
\begin{equation}
\text{EW(\ion{Mg}{2})}_\mathrm{c} = \text{EW(\ion{Mg}{2})} \left(\frac{\langle L_{\text{bol}}/L_{\text{Edd}} \rangle}{L_{\text{bol}}/L_{\text{Edd}}}\right)^{a},
\end{equation}
where the median value of their measured Eddington ratios is $\langle L_{\text{bol}}/L_{\text{Edd}} \rangle$ = 10$^{-0.55}$, and the power-law index is $a=-0.30$. 
We apply this correction to previous samples (except the XQz5 sample; see below) that will be used to compare with our results.

\begin{figure}
%\plotone{FeIIMgII_Edd_r.pdf}
\includegraphics[width=0.48\textwidth]{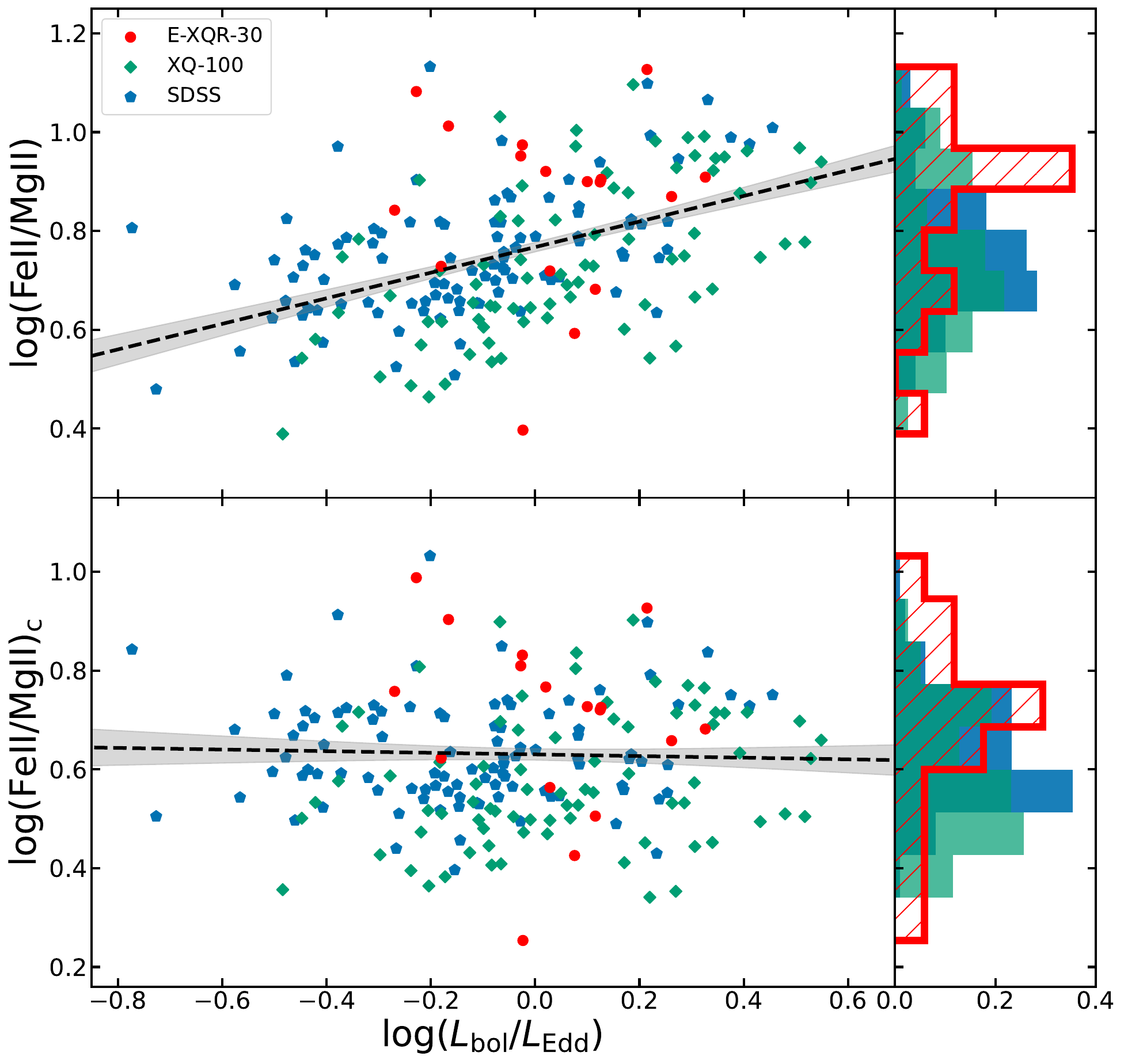}
\caption{Correlation between the Eddington ratio and \ion{Fe}{2}/\ion{Mg}{2} line flux ratio. The upper panel shows the observed correlation and the lower panel shows the correlation after correcting for the dependence of the Eddington ratio (Section~\ref{subsec:EddRatio_correlation}). The symbols and colors are the same as in Figure.~\ref{fig:L_bol_Z}. The side panels show the normalized histograms.
The dashed lines are the linear regressions of the shown data points, and the results are log(\ion{Fe}{2}/\ion{Mg}{2})$=0.26\times$log($L_\mathrm{bol}/L_\mathrm{Edd}$)$+0.77$ in the upper panel and log(\ion{Fe}{2}/\ion{Mg}{2})$_\mathrm{c}=-0.02\times$log($L_\mathrm{bol}/L_\mathrm{Edd}$)$+0.64$ in the lower panel. The uncertainty ranges of the linear regressions  are obtained by the least-square method, and shown as the shaded regions. The weak negative slope of corrected \ion{Fe}{2}/\ion{Mg}{2} in the lower panel is insignificant.
 \label{fig:FeIIMgII_Edd_r}}
\end{figure}

For our samples (i.e., the samples for which we measured the \ion{Fe}{2}/\ion{Mg}{2} ratio ourselves), we slightly modify the correction, because our spectral fitting windows are different from those of  \cite{2017ApJ...834...203} and thus there is an offset between the two \ion{Fe}{2}/\ion{Mg}{2} measurements (see Section \ref{subsec:systematic uncertainty}). We perform a linear regression for the 1291 high-luminosity low-redshift SDSS DR16Q quasars  to derive a new empirical correlation between the Eddington ratio and the \ion{Fe}{2}/\ion{Mg}{2} ratio. From our analyses, the median value of the Eddington ratios is $\langle L_{\text{bol}}/L_{\text{Edd}} \rangle$ = 10$^{-0.62}$ and the power-law index is $a=-0.24\pm0.01$. We propagate this uncertainty of the power-law index into the uncertainty estimate for (\ion{Fe}{2}/\ion{Mg}{2})$_\mathrm{c}$. We use this new correction in our following analyses, assuming that the Eddington ratio dependence does not evolve with redshift. Figure~\ref{fig:FeIIMgII_Edd_r} shows the observed \ion{Fe}{2}/\ion{Mg}{2} ratios before and after this correction, as well as their linear regressions.
We also apply our correction to the XQz5 sample that consists of 83 luminous quasars between $4.5<z<5.3$ with optical and near-IR spectroscopic observations \citep{lai2023a}. The luminosity range of this sample is similar to ours. The \ion{Fe}{2}/\ion{Mg}{2} values of the XQz5 quasars were measured by \cite{lai2023}, and were further corrected to match our measurement for a fair comparison.

\subsection{No Redshift Evolution of \ion{Fe}{2}/\ion{Mg}{2}} \label{subsec:FeIIMgII evolution}

\begin{figure*}
\plotone{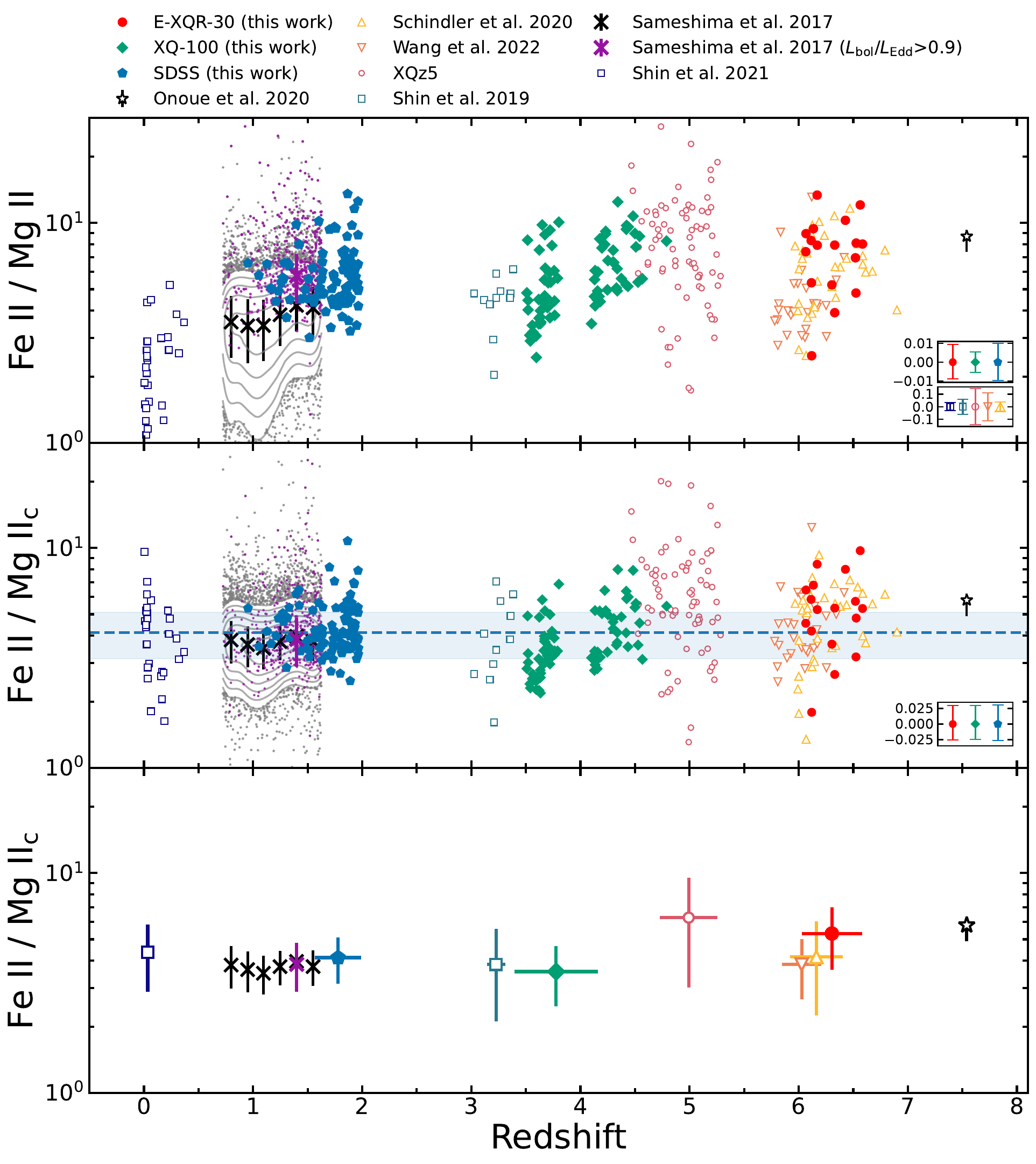}
\caption{\ion{Fe}{2}/\ion{Mg}{2} line flux ratios at redshift from $z=0$ to $7$. The top panel shows the observed flux ratios. The middle panel shows the same ratios after correcting the Eddington ratio dependence (see details in Section~\ref{subsec:EddRatio_correlation}).
The bottom panel shows the median and $\pm1\sigma$ scatter of each sample. The filled symbols show the samples analyzed in this work (red: E-XQR-30, green: XQ-100, blue: low-redshift SDSS), and the open symbols show the samples in the literature. The different colors and symbols are described in the legend on the top. The sub-panels in the lower right corners of the top and middle panels show the typical measurement errors in the logarithmic scale. The grey contours (with a step of 0.2 dex in the logarithmic scale) and grey dots represent the low-redshift sample of \citet{2017ApJ...834...203}. Their median \ion{Fe}{2}/\ion{Mg}{2} flux ratios at six redshift bins (a step of $\Delta z=0.15$) are shown as the black crosses. Those with Eddington ratios over 0.9 are shown as the purple dots, and the median of this subsample is shown as the purple cross. The dashed line and its shaded region are the median value of the SDSS sample and its $\pm1\sigma$ range as our baseline at low redshift. The figure shows no apparent evolution of the line flux ratio with redshift.
\label{fig:FeIIMgII_z}}
\end{figure*}

The \ion{Fe}{2}/\ion{Mg}{2} flux ratios of our three samples are presented in the top panel of Figure~\ref{fig:FeIIMgII_z}.
Other measurements that used the T06 iron template and reported \ion{Fe}{2} flux at the same wavelength range (2200 \AA $\leq$ $\lambda_{\rm rest}$ $\leq$ 3090 \AA) are also shown in the same figure \citep{2017ApJ...834...203, 2019ApJ...874...22, 2021ApJ...917...107, 2020ApJ...905...51, 2020ApJ...898...105, 2022ApJ...925...121}. The middle panel of Figure \ref{fig:FeIIMgII_z} shows the \ion{Fe}{2}/\ion{Mg}{2} line ratios after correcting the Eddington ratio dependence. We show the median and NMAD of the corrected \ion{Fe}{2}/\ion{Mg}{2} ratios for each sample in the bottom panel of Figure \ref{fig:FeIIMgII_z}, and compare their distributions at redshift from $z=0$ to $z>7$. 
We analyze the evolutionary trend of \ion{Fe}{2}/\ion{Mg}{2} using both original \ion{Fe}{2}/\ion{Mg}{2} measurements and corrected values (\ion{Fe}{2}/\ion{Mg}{2})$_{\mathrm{c}}$ from our SDSS, XQ-100, and E-XQR-30 samples. The linear regression shows the slopes of  \ion{Fe}{2}/\ion{Mg}{2} versus redshift are 0.035$\pm$0.034 dex and 0.021$\pm$0.041 dex in the logarithmic scale, respectively. Both slopes are consistent with zero within $1\sigma$.

The median flux ratios before and after the Eddington ratio correction for the E-XQR-30, XQ-100, and SDSS samples are presented in Table~\ref{tab:statistical_FeIIMgII}. We find no significant redshift evolution in the corrected \ion{Fe}{2}/\ion{Mg}{2} flux ratios from $z=0.95$ to 6.59, suggesting early iron enrichment in the vicinity of massive black holes.
Most measurements in the literature also fall within the $1\sigma$ range of the low-redshift SDSS sample. Possible systematic biases in these measurements will be discussed in Section \ref{subsec:systematic uncertainty}. The large scatter of \ion{Fe}{2}/\ion{Mg}{2} flux ratios in the high-redshift sample will be discussed in Section \ref{subsec:larger scatter}.

We also estimate the dependence of the EWs of \ion{Fe}{2} and \ion{Mg}{2} on redshift using our three samples. With the linear regression, we calculate the slopes of the relations between EW and redshift for the two emission lines. The results in the logarithmic scale are $0.027\pm0.065$ and $0.011\pm0.045$, respectively, suggesting no significant evolution within $1\sigma$.

\begin{deluxetable}{lccc}
\tablecaption{\ion{Fe}{2}/\ion{Mg}{2} line ratios of the three main samples\label{tab:statistical_FeIIMgII}}
\tablewidth{0pt}
\tablehead{
\colhead{Sample} & \colhead{Redshift} & \colhead{\ion{Fe}{2}/\ion{Mg}{2}} & \colhead{(\ion{Fe}{2}/\ion{Mg}{2})$_\mathrm{c}$}
}
\startdata
SDSS & $0.95-1.97$ & $5.56\pm1.50$ & $4.13\pm0.98$\\
XQ-100 & $3.51-4.79$ & $5.30\pm1.94$ & $3.57\pm1.09$ \\
E-XQR-30 & $6.05-6.59$ & $7.94\pm2.19$ & $5.31\pm1.68$ \\
\enddata
\tablecomments{The redshifts are based on the \ion{Mg}{2} emission line measurements. For each sample, we show the median value and NMAD as the uncertainty.
}
\end{deluxetable}

\section{Discussion} \label{sec:discussion}

\subsection{Systematic Uncertainties of the \ion{Fe}{2}/\ion{Mg}{2} Measurements} \label{subsec:systematic uncertainty} 

\begin{deluxetable*}{lccccccc}[!]
\tablecaption{Examples of the \ion{Fe}{2}/\ion{Mg}{2} and Eddington ratio results using the T06 and VW01 iron templates\label{tab:t06_vw01_FeIIMgII}}
\tablewidth{0pt}
\tablehead{
\colhead{Name} & \twocolhead{\ion{Fe}{2}/\ion{Mg}{2}} & \twocolhead{$L_\mathrm{bol}/L_\mathrm{Edd}$} & \twocolhead{(\ion{Fe}{2}/\ion{Mg}{2})$_\mathrm{c}$}  &
\\
 \cline{2-7}
 \colhead{}& \colhead{T06} & \colhead{VW01} & \colhead{T06}& \colhead{VW01}& \colhead{T06} & \colhead{VW01} 
}
\startdata
SDSS J0842+1218 & $8.95^{+0.11}_{-0.10}$ & $5.72^{+0.06}_{-0.05}$ & $0.94^{+0.22}_{-0.22}$ & $0.68^{+0.16}_{-0.14}$ &$6.45^{+0.42}_{-0.35}$ & $4.45^{+0.27}_{-0.23}$\\
PSO J158-14 & $7.41^{+0.13}_{-0.12}$ & $5.60^{+0.09}_{-0.08}$ & $1.83^{+0.44}_{-0.42}$ & $1.41^{+0.35}_{-0.37}$ &$4.55^{+0.33}_{-0.26}$ & $3.66^{+0.26}_{-0.22}$ \\
PSO J239-07 & $8.33^{+0.09}_{-0.09}$ & $5.23^{+0.05}_{-0.06}$ & $1.05^{+0.27}_{-0.24}$ & $0.76^{+0.21}_{-0.17}$ &$5.84^{+0.42}_{-0.31}$ & $3.97^{+0.26}_{-0.23}$ \\
\cline{0-7}
Average & 8.23$\pm$0.63 & 5.51$\pm$0.21 & 1.27$\pm$0.40 & 0.95$\pm$0.32 & 5.61$\pm$0.79 & 4.03$\pm$0.32 \\
\enddata
\tablecomments{Each property shows the results measured by both T06 and VW01 iron templates. The last row shows the average and standard deviation of the three quasars results.}
\end{deluxetable*}

Systematic uncertainties of the \ion{Fe}{2}/\ion{Mg}{2} line ratio measurements were not included in the reported uncertainties. They can be caused by different iron templates, spectral fitting windows, emission line fitting models, data reduction, etc. In this section, we discuss some systematic uncertainties of the measurements and their implications. We mostly focus on (\ion{Fe}{2}/\ion{Mg}{2})$_\mathrm{c}$, which is the measured \ion{Fe}{2}/\ion{Mg}{2} corrected for the Eddington ratio dependence using the method in Section \ref{subsec:EddRatio_correlation}.

First of all, the \ion{Fe}{2}/\ion{Mg}{2} measurements rely on iron templates. Our study used the T06 iron template. To test the systematic bias between the T06 and VW01 iron templates, we perform the same fitting method for three good E-XQR-30 spectra using the VW01 template and show the comparison in Table \ref{tab:t06_vw01_FeIIMgII}. 
Compared with the results from the T06 template, the \ion{Fe}{2}/\ion{Mg}{2} line ratios based on the VW01 template are lower by 33$\%$ at a 4 $\sigma$ significance. This is consistent with the results in the literature \citep[e.g.,][]{woo2018, 2019ApJ...874...22, 2020ApJ...905...51, 2020ApJ...898...105}. The (\ion{Fe}{2}/\ion{Mg}{2})$_\mathrm{c}$ values are lower by 28$\%$ at a 2 $\sigma$ significance.

In addition, different fitting windows introduce systemic uncertainties of the measurements. The fitting windows used in our study are 2100--2750 \AA\ and 2850--3100 \AA\ for the continuum$+$iron fitting and 2750--2850 \AA\ for the \ion{Mg}{2} fitting, while \cite{2017ApJ...834...203} used 2200--2700 \AA\ and 2900--3500 \AA\ for the continuum$+$iron fitting and 2700--2900 \AA\ for the \ion{Mg}{2} fitting. In order to check the influence of different fitting windows, we perform the following tests using the SDSS DR16 quasar spectra. We randomly select $\sim500$ spectra and ensure that the rest-frame 2100--3100 \AA\ and 2200--3500 \AA\ spectra of this sample are available. We then fit our model to this sample in the above two different types of fitting windows. The results show that the continuum and the \ion{Fe}{2} and \ion{Mg}{2} line flux measurements are all affected. The \ion{Fe}{2}/\ion{Mg}{2} flux ratios measured using our fitting windows are larger than those using the \cite{2017ApJ...834...203} fitting windows by $\sim15\%$ at a 0.65 $\sigma$ significance. 

Furthermore, different fitting models may affect the \ion{Fe}{2}/\ion{Mg}{2} measurements. 
For example our \ion{Fe}{2}/\ion{Mg}{2} measurements of the XQ-100 quasars are about 30$\%$ larger overall than the \cite{lai2023} results at 1 $\sigma$ significance. We and \cite{lai2023} both measured the XQ-100 sample, but using slightly different fitting methods. We used a single-component Gaussian model to fit the \ion{Mg}{2} emission line, while \cite{lai2023} used a multi-component Gaussian model. We compare the median \ion{Fe}{2}/\ion{Mg}{2} values measured in both works and estimate the systematic bias introduced by the different models. We find a systematic difference of 30$\%$. This is because the multi-component model includes a wider component for \ion{Mg}{2}, and thus provides a lower \ion{Fe}{2}/\ion{Mg}{2} value. This is also mentioned in \cite{2021ApJ...917...107}. When we compare the two results in Figure \ref{fig:FeIIMgII_z}, we have applied a correction to the XQz5 sample to take into account the above systematic difference.

Finally, the spectra of the same quasar observed by different instruments or reduced by different pipelines can be inconsistent and introduce biases. 
We check quasars in common in different samples or studies. SDSS J0842+1218 and PSO J060+24 in our E-XQR-30 sample were included by \cite{2022ApJ...925...121}. We fit their GNIRS near-IR spectra to calculate \ion{Fe}{2}/\ion{Mg}{2} using our fitting method. The calculated \ion{Fe}{2}/\ion{Mg}{2} values are smaller than our results by $\sim 45\%$. This difference is likely due to the low spectral resolution (R $\sim600$) of the GNIRS data \citep{shen2019}. We find that some peaks of the \ion{Fe}{2} emission lines in the low-resolution spectra are smoothed, leading to a smaller value of the \ion{Fe}{2} emission during the modeling process. Therefore, the \ion{Fe}{2}/\ion{Mg}{2} ratios are underestimated. This finding highlights the importance of a relatively high spectral resolution for the measurement of the \ion{Fe}{2}/\ion{Mg}{2} ratio.

\subsection{Robustness of the continuum fitting} \label{subsec:continuum robustness}

In this paper, our measurements are limited to a relatively short wavelength window compared to some previous studies of low-redshift quasars \citep[e.g.,][]{2003ApJ....596...817}. We used the spectra of the XSHOOTER NIR arm, because one single power-law continuum cannot well characterize the entire wavelength range of the XSHOOTER VIS and NIR arms. The truncation at the short wavelength is chosen to be 2100 \AA. This helps us avoid the flux calibration problem in the arm connection range mentioned earlier and thus achieve a good model fit. The truncation at the long wavelength is at 3100 \AA~, which is limited by the effective wavelength coverage of the E-XQR-30 and XQ-100 spectra. In this case, our wavelength window does not cover the Balmer continuum edge $\lambda_{\mathrm{BE}}=$ 3646 \AA, so we have used a fixed Balmer-to-power-law ratio (i.e., $F_{\mathrm{BE}}/F_{\mathrm{PL}}$). In this subsection, we address how our \ion{Fe}{2}/\ion{Mg}{2} measurements were affected by the truncated wavelength windows.

From the SDSS DR16 quasar sample, we select $\sim$200 spectra that cover the following wavelength windows: $\lambda_{\mathrm{rest}}=$ 2100--2750 \AA, 2850--3600 \AA, 4180--4230 \AA~ and 4440--4460 \AA. The last two windows are the two reddest ranges in the SDSS spectra that are dominated by the continuum emission without broad line emission. We perform our spectral fitting method to our truncated wavelength windows and to the new longer windows, respectively. The Balmer-to-power-law ratio is fixed to be 0.3. 
An example of SDSS quasar spectrum with both fitted models is presented in Appendix \ref{sec: truncated windows appendix}.
Figure \ref{fig:windows_comp} shows the comparison of the measured power-law slopes and \ion{Fe}{2}/\ion{Mg}{2} ratios between the two different window ranges. The median slope value (with its NMAD) is $-1.55\pm0.49$ from our truncated wavelength windows, and is $-1.62\pm0.44$ from the longer wavelength windows. The difference is less than $5\%$, suggesting that our truncated wavelength range works reasonably well. The \ion{Fe}{2}/\ion{Mg}{2} ratios are lower when the longer wavelength windows are used, and the difference is 20$\%$ at a $<1\sigma$ significance. We emphasize that this difference does not affect our results, since we use the same truncated wavelength range for all quasars in this study.

We further test the impact of the relative strength of the Balmer continuum using a series of Balmer-to-power-law ratios:  0.1, 0.3, 0.5, 0.8, and 1. We perform our spectral fitting method to our truncated wavelength windows and to the new longer windows, as we did above. Note that the truncated windows do not cover the Balmer edge $\lambda_{\mathrm{BE}}=$ 3646 \AA. For the truncated windows, we fixed the Balmer-to-power-law ratio to be 0.3 earlier. Now we find that different ratios give similar \ion{Fe}{2}/\ion{Mg}{2} measurements within $1\sigma$ errors, which is consistent with the literature results \citep[e.g.,][]{2011ApJ...739...56}. For the longer fitting windows, a Balmer-to-power-law ratio of 0.1 or 0.3 gives the best fit. The best-fit \ion{Fe}{2}/\ion{Mg}{2} measurements are well consistent with those based on the truncated fitting windows. The longer fitting windows still lead to lower \ion{Fe}{2}/\ion{Mg}{2} measurements. The difference is only 10$\%$ at a 0.6$\sigma$ significance. Therefore, it is reasonable to fix the Balmer-to-power-law ratio to be 0.3 for our main samples.

The above tests show that the choice of our spectral fitting windows has not brought obvious systematic biases. Our \ion{Fe}{2}/\ion{Mg}{2} measurements are reliable compared to those based on longer wavelength windows.

\begin{figure}
\includegraphics[width=0.48\textwidth]{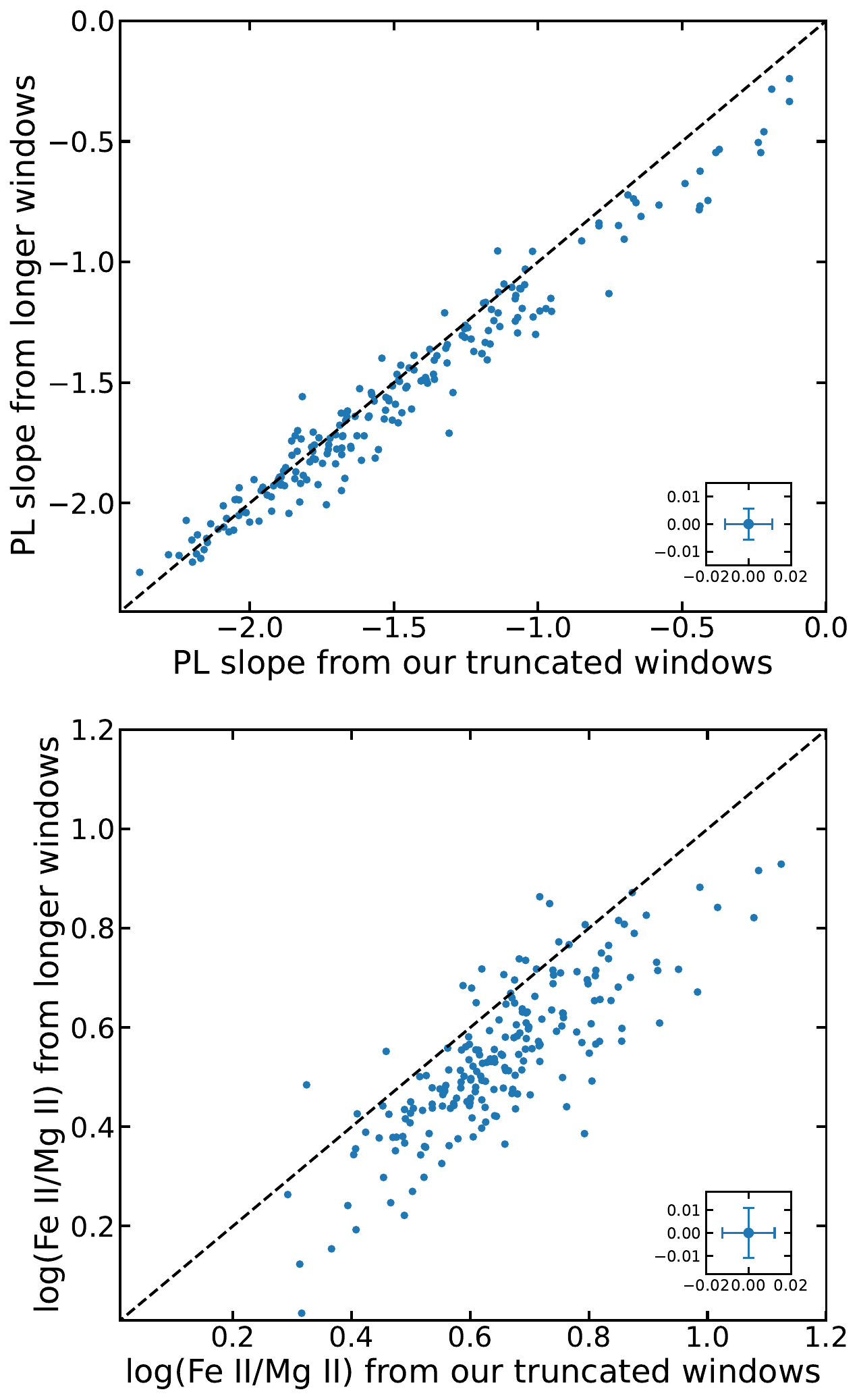}
\caption{
Comparison between our truncated wavelength windows and the longer wavelength windows. The upper and lower panels show the calculated power-law (PL) slopes and \ion{Fe}{2}/\ion{Mg}{2}, respectively. The blue symbols are the measurements of $\sim$200 SDSS DR16 quasars. The black dashed lines indicate the 1:1 reference line. The sub-panel in the lower right corner of each panel shows the typical measurement errors. The Balmer-to-power-law ratio is fixed to be 0.3.
 \label{fig:windows_comp}}
\end{figure}

\subsection{Large scatter of \ion{Fe}{2}/\ion{Mg}{2} in the high-redshift sample} \label{subsec:larger scatter}

We find that the scatter of the \ion{Fe}{2}/\ion{Mg}{2} flux ratios for the E-XQR-30 sample is larger than those for the XQ-100 and SDSS samples, as shown in Table \ref{tab:statistical_FeIIMgII} and Figure 6. 
We have not fully understood the reason for the relatively large scatter at $z>6$, but we have a couple of explanations, including physical and non-physical reasons.

In our quasar selection procedure from the E-XQR-30 sample, we have removed quasars at $z<6.05$ whose spectra are severely affected by the atmospheric absorption in the near-IR. For the remaining quasars at $z>6.05$, we visually checked their spectra and found that their emission line profiles and best-fit models all look reliable. We particularly inspect two quasars with the maximum and minimum corrected \ion{Fe}{2}/\ion{Mg}{2} ratios in this sample, including  PSO~J231-20 ((\ion{Fe}{2}/\ion{Mg}{2})$\rm_c$=9.73$^{+0.76}_{-0.49}$; log($L_\mathrm{bol}$[erg s$^{-1}$])$=47.26$) and CFHQS~J1509-1749 ((\ion{Fe}{2}/\ion{Mg}{2})$\rm_c$=1.79$^{+0.17}_{-0.11}$; log($L_\mathrm{bol}$[erg s$^{-1}$])$=47.34$), and find no apparent anomaly in their \ion{Mg}{2} line profiles. However, we cannot rule out the possibility that some spectra (particularly at $z=6.15$--$6.20$ and $z=6.30$--$6.40$) could have been slightly affected by relatively weak atmospheric absorption at 2.00--2.02 \micron~and 2.05--2.08 \micron. 
 
The large scatter in the high-redshift sample can be attributed to a physical origin. Based on ULAS~J1342+0928 at $z=7.5$, \citet{2022ApJ...937...61} raised a possibility that pair-instability supernovae (PISNe) that emerge from metal-free Population-III stars can supply iron to the BLR cloud in some quasars at high redshift. When PISNe were present, this extra iron supplier may affect the Fe/$\alpha$ abundance ratio. We will further discuss the early iron enrichment in Section~\ref{subsec:iron enrichment}.

\subsection{Iron Enrichment in the Early Universe} \label{subsec:iron enrichment}

Our results confirm the lack of a redshift evolution in the quasar \ion{Fe}{2}/\ion{Mg}{2} line ratios at redshifts up to $z\sim6.6$. 
Thanks to the high data quality of our three samples over a wide redshift range, our measurements provide one of the most robust results for this conclusion. The non-evolution is consistent with the lack of redshift evolution of the BLR metallicity derived from other tracers. Although we do not cover the higher redshift range, the non-evolution trend seems to hold up to $z=7.54$ \citep{2020ApJ...898...105, 2022ApJ...937...61}. The age of the Universe was only 0.8 Gyr at $z\sim6.6$, which is younger than the expected timescale of an efficient iron production by SNe Ia.

The origin for the lack of the \ion{Fe}{2}/\ion{Mg}{2} evolution is still unclear, and some possible solutions have been proposed. The typical time delay between the iron enrichment caused by SNe Ia and SNe II is thought to be about 1 Gyr, but recent observational constraints suggest that the SNe Ia rate has a delay-time distribution (DTD) that has a power-law shape $t^{-1}$ over $t\sim0.1-10$ Gyr \citep{2012PASA...29...447, 2014ARA&A...52...107, 2014AJ...148...13, MaozGraur2017}. Therefore, the cosmic time for the SNe Ia iron enrichment may be earlier and the timescale for the enrichment completion may be statistically shorter than the previous expectation. Many Fe/$\alpha$ evolution models based on SNe Ia DTD and core-collapse supernova (CCSNe) point to a decline in Fe/$\alpha$ at high redshift, but the fact that our \ion{Fe}{2}/\ion{Mg}{2} measurements remain consistent up to $z\sim6.6$ does not support the predictions of these models \citep[e.g.,][]{MaozGraur2017, 2017ApJ...834...203, 2022ApJ...937...61}.

Meanwhile, PISNe are candidates with a significant contribution to the iron enrichment, which is produced by the explosion of Population-III stars as massive as $150-300$ $M_{\sun}$ \citep{2002ApJ...565...385}. The iron yield from one PISNe with 280 $M_{\sun}$ is about one hundred times of that from a SN Ia \citep{2022ApJ...937...61}. Recently, \cite{2023Nature...1} reported a  chemical signature for the existence of PISNe from stars more massive than 140 $M_{\sun}$, highlighting the possibility of the PISNe contribution. Many recent studies also proposed a possible explanation of a rapid iron enrichment by a top-heavy initial mass function (IMF) with a power-law index larger than the classic value, indicating massive CCSNe and PISNe \citep{2022MNRAS...512...2573, ali-dib2023, huang2023, wang2023}. 
More observations and future studies are needed to confirm these candidates.

\section{Summary} \label{sec:summary}

We have analyzed a sample of high-luminosity quasar spectra at $z\sim 6.0-6.6$, focusing on the \ion{Fe}{2}/\ion{Mg}{2} line ratios. This high-redshift sample is from the enlarged XQR-30 project that provided high-quality spectra observed by the XSHOOTER spectrograph at VLT. To make a consistent comparison at different redshifts, we have also analyzed two lower-redshift quasar samples with the same method, including a sample of 78 quasars at $z\sim 3.5-4.8$ from the XQ-100 project observed by VLT XSHOOTER and a sample of 99 quasars at $z\sim 0.95-1.97$ selected from the SDSS DR16Q.

We have performed comprehensive analyses for these luminosity-matched samples and measured their \ion{Fe}{2}/\ion{Mg}{2} line ratios with a correction for the Eddington ratio dependence. 
With a wide redshift coverage and high quality data, our measurements are very suitable to trace the cosmic evolution of \ion{Fe}{2}/\ion{Mg}{2} in a unbiased manner.
Our results confirmed no significant redshift evolution in \ion{Fe}{2}/\ion{Mg}{2} at redshifts from $z\sim1$ to 6.6, suggesting a rapid iron enrichment in the early epoch. The non-evolution of \ion{Fe}{2}/\ion{Mg}{2} can be explained by a time-dependent SNe Ia rate, a non-standard SN population such as PISNe, or a population with a top-heavy IMF. 

The results presented in this paper will motivate further observations of quasars at $z\gtrsim 7$. The first-year JWST observations have discovered a sizable amount of low-luminosity AGNs at high redshift \citep[e.g.,][]{Uebler23, Onoue23, Kocevski23, Harikane23, Matthee23, Maiolino23c}. Follow-up spectroscopy of these new AGN with JWST or future large telescopes will provide an opportunity to identify active SMBHs residing in metal-poor systems. Future near-IR wide-field surveys such as Euclid and Nancy Roman Space Telescope will provide $z\sim7$--$9$ quasars in a similar luminosity range as those presented in this study \citep{Fan19, Euclid19, Tee23}.

\begin{acknowledgments}

We acknowledge support from the National Key R\&D Program of China (2022YFF0503401), the National Science Foundation of China (12225301, 12150410307), and the China Manned Space Project with No. CMS-CSST-2021-A05 and CMS-CSST-2021-A06. 
Y.Z. was supported by the National Science Foundation through award SOSPADA-029 from the NRAO and through grant AST-1751404. R.D. is supported by the Australian Research Council Centre of Excellence for All Sky Astrophysics in 3 Dimensions (ASTRO 3D), through project number CE170100013. E.P.F. is supported by the international Gemini Observatory, a program of NSF’s NOIRLab, which is managed by the Association of Universities for Research in Astronomy (AURA) under a cooperative agreement with the National Science Foundation, on behalf of the Gemini partnership of Argentina, Brazil, Canada, Chile, the Republic of Korea, and the United States of America.

We thank H. Sameshima for providing us his \ion{Fe}{2}/\ion{Mg}{2} data on the SDSS DR7 quasars. We thank L. C. Ho for help comments and suggestions. The data presented in this paper are based on observations collected at the European Southern Observatory under ESO programs 1103.A-0817(A), 189.A-0424(A), 189.A-0424(B).

Funding for the Sloan Digital Sky Survey IV has been provided by the Alfred P. Sloan Foundation, the U.S. Department of Energy Office of 
Science, and the Participating Institutions. SDSS-IV acknowledges support and resources from the Center for High Performance Computing  at the University of Utah. The SDSS website is www.sdss4.org.

\end{acknowledgments}

\facilities{VLT (XSHOOTER), SDSS}
\software{Astropy \citep{2018AJ....156..123A}, Numpy \citep{harris2020array}, Scipy \citep{2020SciPy-NMeth}, Matplotlib \citep{Hunter2007}}

\bibliography{ms}
\bibliographystyle{aasjournal}

\clearpage

\appendix

\section{The degraded E-XQR-30}\label{sec:degraded E-XQR-30 appendix}
\begin{figure}
\centering
 \includegraphics[width=0.6\linewidth]{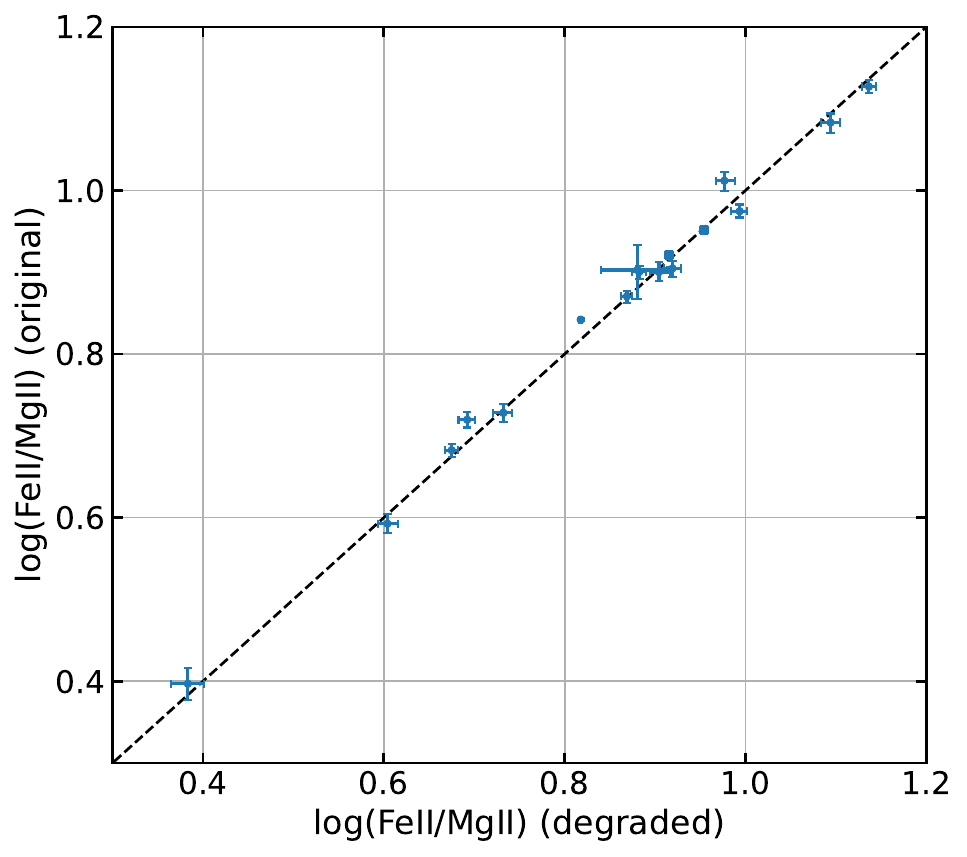}
\figcaption{Comparison of the \ion{Fe}{2}/\ion{Mg}{2} line ratios between the degraded and original E-XQR-30 quasar spectra. The dashed line indicates the 1:1 reference line. \label{fig:degraded_EXQR30}}
\end{figure}

We degraded the E-XQR-30 quasar spectra from their effective spectral resolution \citep{DOdorico2023} to the SDSS spectral resolution R $\sim2000$. Figure \ref{fig:degraded_EXQR30} shows the comparison of the measured \ion{Fe}{2}/\ion{Mg}{2} values of the degraded and original spectra.

\section{Effects of the truncated wavelength windows on spectral fitting} \label{sec: truncated windows appendix}
\begin{figure}
\centering
 \includegraphics[width=0.7\linewidth]{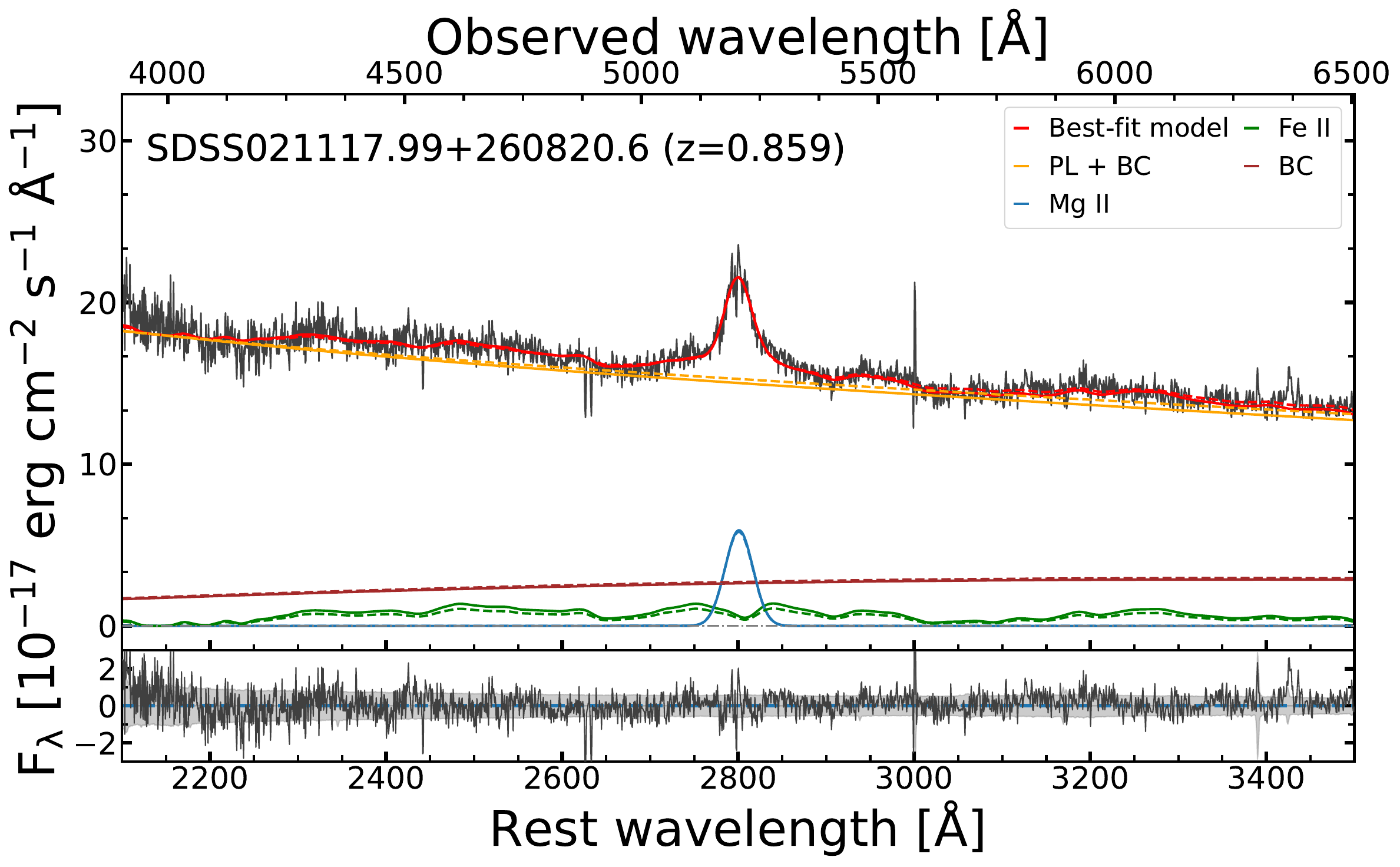}
\figcaption{The color-coded lines have the same meaning as in Figure \ref{fig:fitted_spec_sdss}. The solid lines and dashed lines show the models fitted over our truncated wavelength windows and the longer windows, respectively. 
Since the best-fit models (red lines) are similar, we only display the residual flux of the best-fit model derived from our truncated wavelength windows in the lower subpanel.} \label{fig:fitted_spec_2wins}
\end{figure}

Figure \ref{fig:fitted_spec_2wins} shows the best-fit models of a SDSS spectrum fitted over our truncated wavelength windows and the longer windows, respectively. The Balmer-to-power-law ratio was fixed to be 0.3. The measured \ion{Fe}{2}/\ion{Mg}{2} line ratios are $3.06^{+0.23}_{-0.11}$ and $2.40^{+0.10}_{-0.14}$ respectively.

\end{document}